\documentstyle[prb,aps,epsf]{revtex}

\begin{document}

\draft

\title {Stripe phase: analytical results for weakly coupled repulsive Hubbard model.}

\author{S.I. Mukhin$^1$ and S.I. Matveenko$^2$}
\address{$^1$Moscow Institute for Steel and Alloys,
Theoretical Physics Dept., Leninskii prospect 4,\\
119991 Moscow, Russia; $^2$Landau Institute for Theoretical Physics,
Kosygina Str. 2, 117940, Moscow, Russia}
\date{\today}
\maketitle

\begin{abstract}
Motivated by the stripe developments in cuprates, we review some analytical
results of our studies of the charge- and spin density modulations (CDW and SDW)
in a weakly coupled one dimensional repulsive electron system on a lattice.
It is shown that close to half filling, in the high temperature regime above the
mean field transition temperature, short range repulsions favor charge density
fluctuations with wave vectors bearing special relations with those of the spin
density fluctuations. In the low temperature regime, not only the wave vectors, but
also the mutual phases of the CDW and SDW become coupled due to a quantum
interference phenomenon, leading to the stripe phase instability in a quasi
one-dimensional repulsive electron system. It is shown that away from half filling
periodic lattice potential causes cooperative condensation of the spin and charge
superlattices. "Switching off" this potential causes vanishing of the stripe order.
The leading spin-charge coupling term in the effective Landau functional
is derived microscopically. Results of the 1D renormalization group (parquet) analysis
away from half filling are also presented, which indicate transient-scale correlations
resembling the mean-field pattern. Farther, the self-consistent solution for the
spin-charge solitonic superstructure in a quasi-one-dimensional electron system is
obtained in the framework of the Hubbard model as a function of hole doping and
temperature. Possible relationship with the stripe phase correlations observed in high
$T_c$ cuprates is discussed.
\end{abstract}


Stripe phases recently observed in doped antiferromagnets (cuprates and nickelates)
\cite{1,1a,1b} attract attention to the problem of multi-mode instabilities in the
interacting electron systems. Numerical mean-field calculations and phenomenological
considerations \cite{2,2a,3,3a} suggest a universality of the spin-charge mode coupling
phenomenon in repulsive electronic systems of different dimensionalities. Hence, analytical
mean-field solutions of the multi-mode ordering in 1D system could be revealing with respect
to the mechanism of the mode coupling. The mean-field solutions would be essentially
dimensional-independent and stabilized in three dimensional model by inter-layer
interaction. This review is aimed at discussing some of such solutions.

Inelastic neutron scattering experiments in doped cuprates and nickelates
reveal pronounced dynamical spin fluctuations centered at the wave vectors
$(1/2\pm\varepsilon,1/2)$ and
$(1/2,1/2\pm\varepsilon)$ (in units of $2\pi/a$, $a$ is the lattice constant).
These are accompanied by dynamical `charge' (crystal lattice)
fluctuations at wave vectors $(\pm 2\varepsilon,0)$ and $(0,\pm 2\varepsilon)$.
The characteristic wave vectors of spin and charge are harmonically
related via the doping dependent parameter $\varepsilon$, indicating that
mode-couplings are important. Similarly, the static stripe phases show the
same harmonic relations between the spin- and charge ordering wave vectors,
thus confirming an idea that spin-charge mode coupling is a necessary condition
for these phases to exist. Recent experiments demonstrate \cite{1a,1b} that at
least in cuprates one has also to deal with the coupling to a pairing mode.
It is proposed \cite{3a} that the stripe phase might be a quantum liquid crystal
state of electrons possessing at the same time a charge-, antiferromagnetic-,
and a superconducting condensates.

Since the stripes have to do with a {\em multi-mode} instability,
the experimental discovery of static stripes and
dynamical stripe fluctuations in cuprates and other doped transition metal
oxides \cite{1,1a,1b} put forward the  following theoretical problem:
a theory characterized by coupled modes must be extracted from a (microscopic)
model of an interacting electron system.

Stripe-like instabilities do show up in a
variety of theoretical approaches, some of them even preceding the experimental
discoveries \cite{2,2a}. However, all of these approaches \cite{2,2a,3,3a} have
in common that they deal with the strong coupling regime at zero temperature,
while they rest entirely on numerical calculations. To succeed in the same direction
analytically, a microscopic model with weakly interacting electron-quasiparticles
has been chosen \cite{0,01,muk}. In such a model, in order to get at a mean-field ordering
transition one limits himself to systems exhibiting near to perfect Fermi
surface nesting characteristics. This actually implies the choice of only
one dimensional (1D) systems.

At first glance, one may argue here that the infrared
fixed point structure of 1D interacting electron systems is well understood
\cite{6a}-\cite{7}. Namely, it is well established that the zero temperature Luttinger
liquid exhibits algebraic long range order in both the spin- and the
charge sectors \cite{6b}. Nevertheless, the evidence is growing \cite{jan}
that existing picture is not yet complete. Namely, it is very probable that close
to half-filling and for relatively weak interactions the ground state is nothing else
than a 1D stripe phase, with a long range order which is destroyed by a marginal quantum
fluctuations of the order parameter. It is actually well established, that on the
classical level the ground state in this regime is a stripe phase, since the mean-field
analysis of Schulz \cite{2a} for 2D stripes rests on the nesting features which relate
these results with (in depth) a 1D effect. Quantum mechanics is involved, as usual, in
admixing of the Goldstone modes of the classical state in the ground state, causing in turn
the algebraic long range order.

This review consists of the four main parts. The first two parts deal with
just the two "main" charge- and spin harmonics coupling, while the third and the fourth
parts allow
for all the rest "obertones" selfconsistently, thus introducing solitonic superstructures
of the electronic charge and spin. Before coming to the body of the derivations, let us
make a short outline of the four parts in a more detailed fashion.

The novelty of the approach used in \cite{0,01} and described below lies in that
instead of the conventional expansion in terms of a (small) coupling
constant between electrons, one expands perturbatively in the coupling
strengths between the different spin-  and charge modes. For this purpose the classical
nature of the dynamics of the collective modes is used, which has a
range of validity in a temperature interval around $T_{MF}$ - the mean-field
transition temperature into the ordered state.
Thus, the interaction strength between the electrons translates into strength
of mode couplings. For relatively weak interactions and "high" temperatures
one expects on physical grounds a {\em hierarchy} of mode couplings.
The dominant coupling is found between the fundamental spin- and charge
modes, just as the $\varepsilon$ spin-  and $2\varepsilon$ charge modes of the cuprates.
Higher order couplings are responsible for the "solitonic appearance" of
the stripe ground state, discussed in the last part of the paper.

Then, considering the coupled spin- and charge fluctuations above $T_{MF}$,
it is possible to integrate out spin (SDW) fluctuations in the presence of the
relevant charge (CDW) fluctuations \cite{0,01}. An effective free energy of the
system does posses deep {\em local} minima at finite CDW amplitudes with the different
wave vectors, but the {\em global} minimum still corresponds to no CDW at all.

In the second part, results of the paper \cite{muk} are presented, which
directly demonstrate appearance of the global minimum of the free energy at finite
CDW (and SDW) amplitude below $T_{MF}$ when the mutual phase of the coupled charge-
and spin modes is locked for a constructive interference to occur. Hence, a {\em quantum
interference mechanism} of the stripe phase ordering proposed in \cite{muk} is described.
In order to carry discussion somewhat beyond the mean-field approach, a new modification
of the 1D ("parquet") renormalization group technique is described, which was first applied
in \cite{muk} to the case of coupled charge and spin fluctuations incommensurate with the
crystal lattice away from half filling. As one would expect on the basis of the well known
results for 1D systems mentioned above \cite{6a}-\cite{7}, there is no four-leg vertex
divergence found away from commensurability point in the CDW or SDW channels alone.
Nevertheless, introduction of the infinitesimal vertices with the new "Umklapp wave
vector" $2\varepsilon$, brought by the incommensurate CDW mode, causes divergence of the
four-leg vertices, thus indicating transient-scale correlations resembling the mean-field
stripe pattern.

The last two parts of the review contain results published in \cite{mm}, as well as the new
ones. These concern the
self-consistent solutions of the Bogoliubov-de Gennes equations for a repulsive 1D Hubbard
Hamiltonian written in the Hartree-Fock approximation at zero- and at a finite temperature.
Using methods of the
finite band potential theory \cite{dubrovin}, the coupled solitonic charge- and spin
superstructures are described analytically in terms of the Jakobi elliptic functions, which,
in turn, represent solutions of the non-linear Schr\"odinger equations. It is shown
that far enough from half filling of the bare single electron band the solitonic
superstructures smoothly evolve into two coupled CDW and SDW harmonics, considered in the
first two parts of the paper. A decrease of the effective mode coupling constant accompanies
this transition.

\section{Two harmonics approximation: charge fluctuations in the classical limit.}

This section is based on the results of the works \cite{0,01}.
Our interest is in the description of the precursor fluctuations in the
metallic state of a one dimensional system, at temperatures above the
mean-field ordering temperature. For a single mode instability in the
weak coupling regime this is well described by RPA.
Upon the approach of the mean-field transition, the fluctuations slow
down and in the vicinity of the transition the characteristic
frequency of the collective fluctuations becomes less than temperature.
It was demonstrated by several authors (see e.g. \cite{3=Weger} and references
therein) that under these circumstances
the time dependence of the collective modes can be omitted and instead of
calculating the full quantum-mechanical trace a classical average suffices.

Let us now consider the Hubbard Hamiltonian with $U>0$, written in the fermionic
second-quantized operators $c_{i,\sigma}$:

\begin{equation}
H = t\sum_{<i,j>\sigma}c_{i,\sigma}^{\dagger} c_{j,\sigma} +
 U \sum_i (\frac{1}{4}\hat{n}_{i}^{2}-(\hat{S}_{i}^{z})^{2}) \; ,
\label{hubbard}
\end{equation}
using $\hat{n}_{i\uparrow}\hat{n}_{i\downarrow}=
\frac{1}{4}\hat{n}_{i}^{2}-(\hat{S}_{i}^{z})^{2}$ where $\hat{n} \equiv
\hat{n}_{\uparrow}+\hat{n}_{\downarrow}$ is the fermion density and
 $\hat{S}^{z}$ the $z$-component of the fermion spin. The interaction
term can be decoupled by the Hubbard-Stratanovich transformation,
\begin{equation}
exp [  -\frac{U}{4}( \hat{n}_{i}^{2}-(\hat{S}_{i}^{z})^{2} ) ] =
{ 1 \over {\pi U} } \int d\rho_i d m_i
exp [ - \frac{1}{U} ( \rho_i^2 + m_i^2 ) + i \rho_i \hat n_i +
2 m_i S^z_i ] \; ,
\label{hubstra}
\end{equation}
introducing two auxiliary fields, describing the collective
charge- ($\rho$) and spin ($m$) fluctuations, respectively.
It is noted that at temperatures larger than the
mean-field transition only spin amplitude fluctuations matter and
difficulties \cite{2a}
related to the apparent violation of the global $SU(2)$ invariance by
Eq. (\ref{hubstra}) are of no importance. Conventionally, one proceeds by
neglecting the
charge sector completely, except for the $q=0$ static component of the
charge mode causing a shift of the thermodynamic potential.
However,  since the coupling of the charge- and spin modes is important
at low temperatures the question arises how to deal with these couplings
in the precursor regime.

Starting with a uniform state, it is not easy to keep track of these
mode couplings perturbatively. We assume that both the charge- and
spin mode condense at the
mean-field transition where the dynamics in both sectors slow down
and the quasi static approximation can be applied to either the
spin- or the charge modes, or both. Since the spin modes dominate
the instability, they should be integrated out as accurately as possible
(using RPA) which leaves the charge modes to be taken as the static
ones. Neglecting subdominant charge-charge mode couplings, the trace in the partition
functional $Z$ can
be taken over independent charge modes with wave vector $\delta_c$ and
amplitude $\rho(\delta_c)$,
\begin{equation}
\rho_{i}=\rho (\delta_c)\cos(\delta_{c}x_{i})
\label{rhodef}
\end{equation}
and the partition sum can be approximated as,
\begin{equation}
Z \simeq \displaystyle\int d\rho(\delta_c) d\delta_{c}
Z (\rho(\delta_c), \delta_c)
\label{newz}
\end{equation}
where $Z (\rho(\delta_c), \delta_c)$ is the partition function
describing the dynamics
in the presence of the static charge density waves:
\begin{eqnarray}
Z (\rho(\delta_c), \delta_c) & = &
\displaystyle \left\{ \displaystyle \int {\cal D} \, m_{i}(\tau)
\displaystyle \exp{\left[-\int_{0}^{\beta}\sum_{i}\frac{m_{i}^{2}(\tau)}{4U}
d\tau\right]}Z_{m} (\rho(\delta_c), \delta_c) \right\} \times \nonumber \\
 &  & \left\{\displaystyle
\int {\cal D} \, m_{i}(\tau)\displaystyle\exp{\left[-\int_{0}^{\beta}
\sum_{i}\frac{m_{i}^{2}(\tau)}{4U}d\tau\right]}\right\}^{-1}\label{zrho}\\
Z_{m} (\rho(\delta_c), \delta_c) & = & Tr \displaystyle
\left\{ \exp{[-\beta H_{o}(\rho(\delta_c), \delta_c)]}T_{\tau}\displaystyle
 \exp{\left(\int_{0}^{\beta}\sum_{i}
[m_{i}(\tau)\hat{S}^{z}_{i}(\tau)]d\tau\displaystyle \right)}
 \right\} \label{zrho1}\\
 H_{o}(\rho(\delta_c), \delta_c) & = & \displaystyle
 -t\sum_{<i,j>,\sigma}{a}_{i,\sigma}^{\dagger} a_{j,\sigma} -
\tilde{\mu}\sum_{i}\hat{n}_{i} +
U\sum_{i}\rho_{i}(\hat{n_{i}}- n)/2 +
\frac{U}{4}\sum_{i}\rho_{i}^{2} \label{hrho}
\end{eqnarray}
where $\rho_i$ is given by Eq. (\ref{rhodef}) and
$n$ is the average number of electrons per site ($\leq 1$), while
the chemical potential has been shifted,
\begin{equation}
\tilde{\mu}=\mu-\frac{U}{2}n\; .
\end{equation}
Summarizing, Eqs. (\ref{rhodef}-\ref{hrho}) are a good approximation when
the following conditions are satisfied simultaneously: (i) the characteristic
frequencies of the charge fluctuations should be less than $k_B T$, (ii)
charge-charge mode couplings can be neglected, (iii) The modes
Eq. (\ref{rhodef}) exhaust the partition sum in so far the collective
charge sector is involved. Conditions (i) and (iii) are both controlled by
the slowing down associated with the mean-field transition. Although close
to the transition it is expected that the fundamental charge- and spin
harmonics dominate the thermodynamics, this is not necessarily
the case on microscopic scales, and condition (ii) limits the validity of
our approach to the weak coupling regime. Self-evidently, because the most
important (charge-spin) mode coupling is treated explicitely, the above is
a considerable improvement over conventional single-mode weak coupling
theory, allowing us to penetrate deeper into the intermediate coupling
regime.

\subsection{Mode coupling by Fermi-surface matching.}

In this section we will further elaborate on the formalism. However,
the central outcome of our analysis can already be qualitatively understood
at this point: pairs of charge- and spin fluctuations with special
relationships between their momenta are {\em simultaneously} enhanced in
the approach to the transition. These include the $\varepsilon$-spin and
$2\varepsilon$ charge fluctuations reminescent of the stripe fluctuations
in cuprates. In addition, there is actually an infinite progression of
these pairs, although the enhancement factors rapidly decrease upon going
to higher orders. The mechanism is straightforward \cite{0}: by repeated Umklapp
scatterings against the CDW, a non-$2k_F$ SDW can `borrow' the momentum
mismatch with regard to the $2k_F$ nesting vector from the CDW. Obviously,
this can only happen if special relations exist between the wave vectors
of the SDW and CDW.

Close to the transition, the spin fluctuation can be considered as static,
as the static charge density wave Eq. (\ref{rhodef}). Due to the underlying
lattice periodicity a spin-density wave is of the form
\begin{equation}
S(x)=\displaystyle S_{0}\cos (\delta_{s}x) \sin (\frac{\pi x}{a})=
S_{o}\frac{1}{2}\left[\sin (x k_{-}) + \sin (x k_{+})\right]
\label{spindef}
\end{equation}
with $k_{\pm}=\pi/a \pm \delta_{s}$.
The two terms in the brackets in Eq. (\ref{spindef}) have a straightforward
meaning. The first one arises due to the back-scattering of the fermion with a
momentum change of $k_{-}$, while the second term comes
from Umklapp scattering with a total momentum change
$2\pi/a-k_{-}\equiv k_{+}$.
On the other hand, assuming linear fermion dispersions the Fermi-momentum
can be written in the proximity of half-filling as,
\begin{equation}
k_{F}=\frac{\pi}{2a}-\frac{\delta}{2}\; ;\; \frac{\delta}{2}
\equiv x_{d}\frac{\pi}{2a}
\label{kfdef}
\end{equation}
where $x_{d}$ is the (hole) doping concentration (at half-filling,
 $k_{F}=\pi/(2a)$ ).

We have introduced intentionally three different momenta: $\delta$, the
deviation of $k_F$ from its half-filled value, and the wavectors associated
with the modulation of the N\'eel state ($\delta_s$) and with the charge
modulation ($\delta_c$), respectively. Only in the absence of the CDW, the
SDW wave vector  $k_{-}$ equals $2k_{F}=\pi/a-\delta\equiv\pi/a-\delta_{s}$.
In the presence of the CDW, however, a  whole variety of new possibilities
arises. According to Eq. (\ref{hrho} ), the fermion momentum is conserved
only {\it{modulo}} $m\times\delta_{c}$, $m=\pm 1,\pm 2$,..., because of
the presence of the periodic potential produced by the CDW.
This adds new umklapp scattering vectors, which are
linear combinations of the vectors $2\pi/a$ and $\delta_{c}$.
Therefore, $2k_{F}$ may now differ from either $k_{-}$ or $k_{+}$ by an integer
number of CDW wave vectors: $k_{-}-2k_{F}= m\delta_{c};\; m=0,\pm 1,...$;
$k_{+}-2k_{F}= n\delta_{c};\; n=0, 1,...$. Notice that the summation over
$n$ is restricted to positive integers to avoid redundancy in the counting
of the allowed SDW wave vectors. These relations are equivalent to,

\begin{eqnarray}
&&\delta-\delta_{s}=m\delta_{c}\; ; \;\;m=0,\pm 1,\pm 2,...\nonumber\\
&&\delta+\delta_{s}=n\delta_{c}\; ; \;\;n= 1, 2,...
\label{match1}
\end{eqnarray}

\noindent
It follows immediately that both relations are only fulfilled
{\it{simultaneously}} for the following "matching pairs" \cite{0} of wave vectors
$\delta_{c}$ and $\delta_{s}$,

\begin{eqnarray}
 \displaystyle \left\{  \begin{array}{c}
 \delta_{c}=\displaystyle\frac{2\delta}{n-m} \\
 \\
\; \; \; \delta_{s}=\displaystyle\frac{(m+n)\delta}{n-m}
\end{array} \right.
\label{match2}
\end{eqnarray}

\noindent
where $|m|<n\; ; m=0,\pm 1,\pm 2,...\; ; n=1,2,...$. The distinction
between the matching pairs in Eq. (\ref{match2}) and other pairs
 $\delta_{c}$ and $\delta_{s}$, obeying only {\it{one}} of
the two relations in Eq. (\ref{match1}) will become apparent in the next subsection.
Notice that the symmetry relations expressed in Eqs.
(\ref{match1}-\ref{match2})
are not related to the actual values of $t$ and $U$ in the Hamiltonian
Eq. (\ref{hubbard}), as long as the fermion dispersions are linear
and the two-mode approximation (one SDW and one CDW) is valid. In combination,
this limits the present approach to relatively weak couplings.
We will elobarate this further in the next sections.

\subsection{Free energy of the SDW-CDW fluctuations}

Let us proceed to derive the fluctuation contribution to the free energy,
in order to see how the relations  Eqs. (\ref{match1},\ref{match2})
follow from the thermodynamics of the electron system described
by Eqs. (\ref{rhodef}-\ref{hrho}). For this purpose, the
spin fluctuations in the presence of the CDW, as described by
Eqs. (\ref{zrho}-\ref{hrho}) will be integrated out on the RPA level.
We proceed by calculating the free energy functional for fixed
charge modulations,
\begin{equation}
\Omega(\rho(\delta_c), \delta_c) = -T \ln Z (\rho(\delta_c), \delta_c)\; ,
\label{omdef}
\end{equation}
with $Z (\rho(\delta_c), \delta_c)$ defined in Eq. (\ref{zrho}) and
the total partition sum Eq. (\ref{newz}) becomes,
\begin{equation}
Z = \int d\delta_c \int d\rho(\delta_z) \exp (
- \beta \Omega(\rho(\delta_c), \delta_c) )
\label{znewdef}
\end{equation}

In the single-loop approximation, $Z_{m} (\rho(\delta_c), \delta_c)$
in Eq. (\ref{zrho1}) becomes

\begin{eqnarray}
Z_{m} (\rho(\delta_c), \delta_c)
& =&\displaystyle\exp\{-\beta(\Omega_{o}+\Omega_{m})\}\; ;\; \Omega_{o}\approx
\frac{U(1-U\nu(\tilde{\mu}))}{4}\sum_{i}\rho_{i}^{2}+\Omega_{b}
\label{rpa1} \\
\displaystyle\beta\Omega_{m}& = &
 -\frac{1}{2}\int_{0}^{\beta}\int_{0}^{\beta}d\,\tau_{1}d\,\tau_{2}
\sum_{i_{1},i_{2}}m_{i_{1}}(\tau_{1})m_{i_{2}}(\tau_{2})\langle T_{\tau}
\hat{S}^{z}_{i_{1}}(\tau_{1})\hat{S}^{z}_{i_{2}}(\tau_{2})\rangle \nonumber\\
&= &\displaystyle -\sum_{\omega_{n},\vec{q}}m(\vec{q},\omega_{n})
m(-\vec{q},-\omega_{n})\chi(\vec{q},\omega_{n})\; , \label{rpa2}
\end{eqnarray}

\noindent
where  $\nu(\tilde{\mu})$ is the density of states at the Fermi level, and
$\chi(\vec{q},\omega_{n})$ the Fourier component of the magnetic
 susceptibility
of the system calculated with the Hamiltonian Eq. (\ref{hrho}), i.e. in the presence of the
charge density modulation $\rho_{i}$, Eq. (\ref{rhodef}).
$m(\vec{q},\omega_{n})$ corresponds with the Fourier component of the
Hubbard-Stratonovich spin field at momentum $q$ and Matsubara
frequency $\omega_{n}$\cite{2c}.
Finally, $\Omega_{b}$ is a background contribution which is
independent of $\rho_{i}$ and $m$.
Substituting Eqs. (\ref{rpa1},\ref{rpa2}) into Eqs. (\ref{zrho},\ref{zrho1})
and carrying out a
Gaussian integration over the real and imaginary parts of the Fourier components
$m(\vec{q},\omega_{n})$ we arrive at the following expression for the free energy
functional Eq.(\ref{omdef}),

\begin{equation}
\Omega(\rho(\delta_c),\delta_c)=
\displaystyle\frac{U(1-U\nu(\tilde{\mu}))}{4}\sum_{i}\rho(\delta_c)^{2}+
\Omega_{b}+TN\sum_{\omega_{n},\vec{q}}\ln{(1-U\chi(\vec{q},\omega_{n}))}
\label{omeg}
\end{equation}

\noindent
Dropping as usually \cite{3=Weger} all the terms with $\omega_{n}\neq 0$,
the semi-static  part of the free energy functional (per lattice site)
$\tilde{\Omega}$ is found, depending on the CDW with wave vector $\delta_c$
and amplitude $\rho(\delta_c)$,

\begin{equation}
\tilde{\Omega}(\rho(\delta_c,\delta_{c})=
\displaystyle \frac{U\rho(\delta_c)^{2}}{4}(1-U\nu(\tilde{\mu}))+
T\sum_{q=\pm\delta_{s}}\ln(1-U\chi_{\pi/a+q}(\rho(\delta_c), \delta_{c}))
+\tilde{\Omega}_{b}\; .
\label{ometil}
\end{equation}

\noindent
$\tilde{\Omega}_{b}$ absorbs the background contributions and
will be neglected in the remainder.
The first term describes the potential energy cost of creating a CDW
in a system with repulsive interactions, and the
second term describes the decrease of the free
energy due to the spin density fluctuations with wave vectors $\delta_s$
relative to the wave vector of the antiferromagnet at half-filling.

The static magnetic susceptibility $\chi$ in the presence of the periodic
CDW potential remains to be calculated. For linearized dispersion of electrons
near Fermi momentum it takes
the form (compare \cite{Gorkov}),
\begin{eqnarray}
\chi_{\pi/a + \delta_{s}}(\rho(\delta_c), \delta_{c})
& = & T\sum_{\omega_{n}}\int d\, x'\,
\left[\exp\{i(\delta_{s}-\delta)(x-x')\}
 g_{++}(i\omega_{n};x,x')g_{--}(i\omega_{n};x',x)\right.\nonumber\\
&&\left. +  (x\leftrightarrow x')\right]\; \label{semchi},
\end{eqnarray}

\noindent
counting $2k_F$ from its commensurate $\pi/a$ value according to Eq.
(\ref{kfdef}).
The Green's function $g_{++}$ ($g_{--}$) is the
 slowly varying part of the full fermion
Green's function, $G$, in the Matsubara representation
for the right (left)-movers. For example,

\begin{equation}
G_{++}(i\omega_{n};x,x')=\exp[ik_{F}(x-x')]g_{++}(i\omega_{n};x,x')\; .
\end{equation}

\noindent
For these Green's functions we find,

\begin{eqnarray}
g_{++}(i\omega_{n};x,x')& = & \displaystyle\frac{sign\omega_{n}}{i2t}
\exp\left[-\frac{\omega_{n}(x-x')}{2t}- \frac{iU}{4t}\int_{x'}^{x}\rho (u)d\,
u\right];\;\mbox{when $\displaystyle\frac{\omega_{n}(x-x')}{2t}>0$}\nonumber \\
g_{++}(i\omega_{n};x,x') & = & 0; \;\mbox{when $\displaystyle\frac{\omega_{n}(x-x')}
{2t}<0$
}\label{gright}
\end{eqnarray}

\noindent
using the shorthand $\rho (u)=\rho(\delta_c) \cos (\delta_{c}u)$ for
the charge modulation. The corresponding Green's function for the
left-movers is derived by changing the sign in front of the $t$
appearing in the argument of the exponent in Eq. (\ref{gright}).
Substituting Eq. (\ref{gright}) into Eq. (\ref{semchi}), and making use of
the identity :

\begin{eqnarray}
\sum_{n=0}^{\infty}\exp{\{-\pi T(2n+1)z\}}\equiv\displaystyle\frac{1}{\sinh{\{\pi
Tz}\}};\;\; z>0\nonumber
\end{eqnarray}

\noindent
one obtains:

\begin{eqnarray}
\chi_{\pi/a +\delta_{s}}&=&\displaystyle\frac{T}{4t^{2}}Re\left(\int_{1}^{\infty}dz\;
\frac{\exp{\left[iz(\delta_{s}-\delta)-\displaystyle\frac{i2U\rho(\delta_{c})}
{\delta_{c}t}\sin{(\delta_{c}z/2)}\cos{(\delta_{c}(x-z/2))}\right]}}{\sinh{(\pi Tz/t)}}
\right. \nonumber \\
&&\displaystyle \left. +\int_{1}^{\infty}dz\;
\frac{\exp{\left[iz(\delta_{s}+\delta)+\displaystyle\frac{i2U\rho(\delta_{c})}
{\delta_{c}t}\sin{(\delta_{c}z/2)}\cos{(\delta_{c}(x-z/2))}\right]}}
{\sinh{(\pi Tz/t)}}  \right)
\label{chi1}
\end{eqnarray}

\noindent
where lower cut-off in the integration is taken at the lattice constant $a \equiv 1$.
Now, using wellknown property of Bessel functions of integer order:

\begin{eqnarray}
\exp{(iz\cos{\phi})}=\sum_{n=-\infty}^{n=+\infty}i^{n}\exp{(in\phi)}J_{n}(z)\nonumber
\end{eqnarray}

\noindent
and  averaging over the phase (position) of the CDW in Eq. (\ref{chi1}) we find:

\begin{eqnarray}
\chi_{\pi/a+\delta_{s}}
=\displaystyle\frac{T}{4t^{2}}\int_{1}^{\infty}dz\;\frac{[\cos{(z(\delta_{s}+\delta))}+
\cos{(z(\delta_{s}-\delta))}]}{\sinh{(\pi
Tz/t)}}J_{0}\left(\frac{2U\rho(\delta_{c})}{\delta_{c}t}\sin{(\delta_{c}z/2)}\right)
\label{chi2}
\end{eqnarray}

\noindent
Using now the well known \cite{5} addition theorem for
 Bessel functions of integer order ($J_{n}(x)$), we obtain the key result:

\begin{eqnarray}
&&\chi_{\pi/a
+\delta_{s}}=\displaystyle\frac{T}{4t^{2}}\int_{1}^{\infty}
\frac{dz}{\sinh{(\pi Tz/t)}}\left[J_{0}^{2}
\left(\frac{U\rho(\delta_c)}{t\delta_{c}}\right)[\cos{(z(\delta_{s}+\delta))}+
\cos{(z(\delta_{s}-\delta))}] \right. \nonumber \\
&&\left. + \sum_{n=1}^{\infty}J_{n}^{2}\left(\frac{U\rho(\delta_c)}
{t\delta_{c}}\right)[\cos{(z(\delta_{s}+\delta+n\delta_{c}))}+
\cos{(z(\delta_{s}-\delta+n\delta_{c}))}+\cos{(z(\delta_{s}+\delta-n\delta_{c}))}
\right.\nonumber \\
&&\left. +\cos{(z(\delta_{s}-\delta-n\delta_{c}))}]\displaystyle \right]
\label{chi3}
\end{eqnarray}

\noindent
Coarse graining of Eq. (\ref{chi3}) with respect to the thermal length of electron,
$l_{T}=\pi T/t$, leads to the simplified expression (\cite{0}):

\begin{eqnarray}
\chi_{\pi/a +\delta_{s}}
(\rho(\delta_c), \delta_{c})& = & \displaystyle\frac{1}{4\pi t}\ln{\frac{2t}{\pi T}}
\left[J_{0}^{2}\left(\frac{U\rho(\delta_c)}
{t\delta_{c}}\right)(\delta_{\delta_{s},\delta}+\delta_{\delta_{s},-\delta}) +
\sum_{n=1}^{\infty}
J_{n}^{2}\left(\frac{U\rho(\delta_c)}{t\delta_{c}}\right) \right. \nonumber \\
   & & \left. \times (
\delta_{\delta_{s} + \delta, n\delta_{c}} +
\delta_{\delta_{s} - \delta, n\delta_{c}} +
\delta_{\delta_{s} + \delta, - n\delta_{c}} +
\delta_{\delta_{s} - \delta, - n\delta_{c}} )
\right]\;   \label{chibes}
\end{eqnarray}

\noindent
where $\delta_{\alpha,\beta}$ is the Kronecker symbol. For a zero amplitude of
the CDW ($\rho(\delta_c) =0$), Eq. (\ref{chibes}) reduces to the
standard weak coupling result\cite{3=Weger} and the SDW condensation
temperature follows directly,

\begin{equation}
T_{SDW}=\displaystyle \frac{2t}{\pi}\exp{\left(-\frac{4\pi t}{U}\right)}
\label{Tsdw}
\end{equation}

The picture changes drastically when the CDW's are allowed to have a
finite amplitude. To see what happens, let us introduce the distribution
function describing the probability of finding a CDW fluctuation at
wave vector $\delta_c$,

\begin{equation}
W(\delta_{c})=\displaystyle\int\exp\left\{-\frac{\tilde{\Omega}
(\rho(\delta_c),\delta_{c})}{T}\right\}d\,\rho(\delta_c)
\left[\displaystyle\int\left(\int
\exp\left\{-\frac{\tilde{\Omega}
(\rho(\delta_c), \delta_{c})}{T}\right\}d\,\rho(\delta_c) \right)d\,
\delta_{c}\right]^{-1}\; ,
\label{dist2D}
\end{equation}

\noindent
which was calculated numerically. The effect of the thermal length, $l_{T}$, is
effectively excluded in the simplified expression in Eq. (\ref{chibes})
(coarse graining procedure) and is present in Eq. (\ref{chi3}).
This will especially be of importance at large wavelengths, $\delta_c
< \pi / l_{T}$, diminishing the multiple scatterings of the fermions against
the CDW. It is noticed that in this way only an upper bound to the
disordering length is incorporated. Correlations are expected to be
further reduced by charge-charge mode couplings, etcetera, neglected
in the present analysis.

In the figures 1-2 the numerical results following from Eq. (\ref{dist2D})
are shown for some representative choices of parameters. Namely, in Fig.1 :
 $x_{d}=0.02,\; U/t=2;\; T=1.2T_{SDW};\; T=2T_{SDW}$.
The blue dashed ($T=1.2T_{SDW}$) and red dashed ($T=2T_{SDW}$) curves are obtained
using Eq. (\ref{chibes}), i.e. when damping
effects have been neglected. The blue solid ($T=1.2T_{SDW}$) and red solid
($T=2T_{SDW}$) curves in Fig. 1 were calculated using Eq. (\ref{chi3}), i.e. with
the thermal length effect of the fermion included. This makes
peaks sensitive to the temperature, so that higher the temperature
the more substantial becomes smearing of the peaks. Comparing dashed and solid
curves we notice that finite thermal length causes merging of the peaks for small
values of $\delta_{c}$ (corresponding to $Z_{c}<1$) into one broad feature, while
these peaks remain well resolved on the dashed curves at both temperatures
chosen.
For reasons which will become clear in a moment, the wave vectors are normalized
to the Fermi surface spanning vector $\delta$,

\begin{eqnarray}
Z_c & \equiv & { {\delta_c} \over {\delta} } \nonumber \\
Z_s & \equiv & { {\delta_s} \over {\delta} } \label{kdef}\; .
\end{eqnarray}

Let us first consider the undamped case in Fig. 1. Our main result
becomes immediately obvious: {\em the charge fluctuations exhibit a
highly organized behavior as function of their wave vectors.}
The large momentum regime is dominated by a peak centered at $Z_c =2$,
corresponding with CDW wave vector $\delta_c = 2 \delta$ which
is reminescent of the stripe-charge fluctuation as seen in cuprates. This
fluctuation becomes more significant both for smalller dopings and smaller
temperatures.\\
In addition, a variety of smaller peaks is found which occur at
{\em finite CDW amplitudes} and at wave vectors which are {\em rational fractions
of the Fermi-surface spanning factor $\delta$}, i.e.
$Z_c \equiv \delta_c/\delta = n / m$, with $n,m$ integers.
Since the
fluctuations under $Z_c =2$ are not completely unanticipated, these finite
amplitude `fractional momentum' charge fluctuations should be viewed as
a qualitative novelty, unique to the present analysis.
Because they show up at smaller momenta, they are also more susceptible
for smearing effects as the comparison with the two solid/dashed curves in Fig. 1
and two curves in Fig. 2 show.
 At the same
time, it might well be that the members of this series with the largest
$Z_c$'s will survive.  Fig. 2 is important for
understanding of the Scilla and Horribda situation created by the first two terms in
 Eq. (\ref{ometil}). While the first term suppresses peaks at too high doping due to
increase of potential ("Coulomb") energy cost of a CDW, the second term produces
less sharper peaks at too small doping due to thermal length smearing effect at
small wave vector value, $\delta_{c}\sim x_{d}$, of a CDW. In order to decrease
smearing effect at a fixed doping one has to go to lower temperatures, i.e. smaller
$T_{SDW}$, by decreasing $U/t$ ratio. Solid (black)curve in Fig. 2 is obtained for the
set of parameters: $x_{d}=0.001,\; U/t=1;\; T=1.2T_{SDW}$. Dotted (red) curve in Fig. 2
is the same as the "low temperature"  (i.e. at $T=1.2T_{SDW}$) solid (blue) curve in
Fig. 1, and is drawn for comparison. We see that peak at $\delta_{c}=\delta$ ($Z_{c}=1$) has
become well resolved,
and both peaks on the solid curve are much sharper than on the dotted one. This means
that {\em effect of CDW precursors with discrete wave vectors demonstarted in Fig. 1
becomes well pronounced in the weak coupling-low doping concentration limit}.

The attentive reader should already have realized that the above reflects
the `matching pair' counting of Section A.
Recalling Eq. (\ref{match2}), the first few members of this series are:
$\{\delta_{s}=\delta\; , \delta_{c}=2\delta\}$;
 $\{\delta_{s}=\delta\; , \delta_{c}=\delta\}$;
$\{\delta_{s}=\delta\; , \delta_{c}=2\delta/3\}$, and
$\{\delta_{s}=\delta \; , \delta_{c}=\delta/2\}$, which
corrrespond with the sets of integers : $m=0; \; n=1,2,3,4$. The charge wave vectors
 are clearly recognized in the figures (except $\delta_{c}=2\delta/3$, which is not
seen due to not sufficient sensitivity of the Kronecker symbol simulation routine at
$Z_{c} = n/3,\; n=1,2$).

It is instructive to consider how these `matching pair' fluctuations arise
in the present calculation. The key is that the CDW fluctuations with
a `matching' wave vector lead to a selective enhancement of the spin
density fluctuations with proper momenta. The latter push the free-energy
downwards, causing pronounced local minima in
$\tilde{\Omega}(\rho(\delta_c,\delta_{c})$ (Eq. \ref{ometil}) which are
sufficiently deep to carry appreciable statistical weight in Eq. (\ref{dist2D}).

The dependence of the free-energy on the CDW wave vector $\delta_c$ enters
entirely through the generalized Lindhard function, Eqs.(\ref{chi3}),(\ref{chibes}).
 The
latter depends on this wave vector both through the argument of the Bessel
functions, and through the limitations imposed on the summation over the
higher order Bessel functions by the Kronecker $\delta$'s, matching the
SDW and CDW vectors to the Fermi momentum. On the other hand, the dependence
on the amplitude $\rho(\delta_c)$ enters both through the first term
in Eq. (\ref{ometil}), describing the restoring force tending to keep the
charge density uniform in this repulsive system, and again through the
argument of the Bessel functions. The first term $\sim J^2_0$ in
Eq. (\ref{chibes}) is non-zero for any CDW wave vector and is responsible
for the $\delta_c$ independent background. The texture in Figs. 1-2 is
caused by the higher order Bessel functions and in order to gain some
intuition one has to investigate
which values of $\delta_{s}$ contribute to the sum in Eq. (\ref{ometil})
at any fixed
value of $\delta_{c}$, as well as how differs the actual structure of the expression
$\chi_{\delta_{s}}$ in Eq. (\ref{chibes}) for the values of $\delta_{c}$, which either
 belong or
 do not belong to any matching pair $\{\delta_{c},\delta_{s}\}$.
 For this purpose the
manifold of points $\{\delta_{c},\delta_{s}\}$, which give non-zero contribution into
$\chi_{\delta_{s}}$ in Eq. (\ref{chibes}), is plotted in Fig. 3 in the coordinates
$Z_{c}$, $Z_{s}$. At each point belonging
 to any straight line in Fig.3 : $Z_{s}\pm 1 = n\times Z_{c};\;
 n=0,\,\pm 1,\,\pm 2\,,...$,
an argument of one of the Kronecker's symbols in Eq. (\ref{chibes}) becomes zero .
 Only Bessel functions
 of the order $n\leq 8$
were retained in the sum in r.h.s. of Eq. (\ref{chibes}) in order to obtain plots
in Figs. 1-2.
 The reason for this
is that high order Bessel functions,
$J_{n}^{2}(U\rho(\delta_{c})/(t\delta_{c}))$, give negligible contributions to the
 sum
 when their argument becomes much less than order $n$. The value of the argument,
$z\equiv U\rho(\delta_c)/(t\delta_{c})$, is limited from above by the condition of the
 high
probability of the related CDW fluctuation:
$\tilde{\Omega}(\rho(\delta_c),\delta_{c})/T\leq 1$.
This condition leads, due to the first
term in Eq. (\ref{ometil}), to the limitation: $z\leq 4\exp{(-w)}/(\pi x_{d}\sqrt{w})$,
 where
$w\equiv 4\pi t/U$, and the estimates: T $\sim$ T$_{SDW}$, $\delta_{c}\sim \delta$ were
 used. For the range of doping concentrations $x_{d}$ and coupling strengths $w$,
which were chosen for the calculations, the last inequality gives $n\, < \, 5$.
 The coordinates of the crossing points of any two lines
in fig. 3 determine the matching pairs of CDW-SDW fluctuations with wave-vectors
$\{\delta_{c}=Z_{c}\delta,\;\delta_{s}=Z_{s}\delta \}$.
In order to see the "matching - non-matching" distinction in the structure of the
expression for $\chi_{\delta_{s}}$ in Eq. (\ref{ometil}), Fig. 4 is presented. It
shows how many terms are simultaneously non-zero in Eq. (\ref{chibes}) at each point,
 which belongs to the
manifold in Fig. 3 and has coordinates $\{Z_{c},\, Z_{s}\}$.
The latter manifold was substantially rarefied along $Z_{c}$
(but not along $Z_{s}$ !) axis in order to provide better visibility in Fig. 4.
 Each solid circle in Fig. 4
indicates that the number of non-zero terms at this particular point is two,
 while open circles and squares indicate that
this number is one.  Again, only Bessel
functions of the order $n\leq 4$ are accounted for. The solid circles in Fig. 4
coincide with the two line crossing
points in Fig. 3. We see that "columns" of solid circles in Fig. 4 erect only upon the
$\delta_{c}$ ($Z_{c}$) values which are members of the matching pairs. The open
circles in these columns would be also solid if we would not restrict the order of
the Bessel functions in Eq. (\ref{chibes}) to $n\leq 8$.
 The structure of the sum in Eq. (\ref{chibes}) described above gives rise
 to the peaks seen in the W$(Z_{c})$ curves in Figs. 1-2. Finally, the absolute walue
of W$(Z_{c})$ is obtained by normalizing to unity on the limited interval of
variation of $Z_{c}$. This interval is limited from above
by the applicability of the quasi classical approximation for the Green's functions
of electron, i.e. $\delta_{c}\ll \pi/a\sim 2k_{F}$, or in the
equivalent form: $Z_{c}\ll 1/x_{d}$. On the other hand,
it is seen in Figs. 1-2 that the background part of W$(Z_{c})$ stretches
over an interval of $Z_{c}$, which is at least order of magnitude wide,
and extrapolates to a non-zero value in the "forbidden region", $Z_{c} \sim 1/x_{d}$.
This makes normalization procedure somewhat uncertain.

\subsection{Discussion: $\delta_{c}=2\delta_{s}$, but $\delta_{s}=x_{d}/2$}

Here the main results presented in  Section I are summarized and discussed.
First of all, we notice that in terms of
 reciprocal units, $2 \pi /a$, the wave vectors of the symmetry coupled SDW and CDW
fluctuations, `matching pairs', chosen at the end of the previous subsection, could be
 expressed as:
 $1/2 \pm \varepsilon$ , $\pm 2\varepsilon$ for the first pair;  $1/2 \pm \varepsilon$ ,
 $\pm \varepsilon$ for the second pair; $1/2 \pm 3\varepsilon$ , $\pm 2 \varepsilon$ for the
 third pair, and $1/2\pm \varepsilon/3$ , $\pm 2 \varepsilon /3$ for the fourth
matching pair of the wave-vectors $\{\delta_{s},\delta_{c}\}$. Here we have
according to Eq. (12): $\varepsilon=a\delta/(2\pi) = x_{d}/2$. The whole picture of the
 CDW fluctuation loses its sense if the period of the wave, $2\pi/\delta_{c}$, becomes
 greater than the thermal length, $l_{T}\sim at/(\pi T)$
 (compare \cite{3=Weger,Gorkov}). Allowing for the fact that the most
important values are $\delta_{c}\sim \delta_{s}\sim\delta$ we find that our results
are limited to the interval of doping concentrations not too close to half-filling:

\begin{equation}
x_{d}\gg\displaystyle\frac{T}{t}\geq\displaystyle\frac{T_{SDW}}{t}\sim
\exp{\left(-\frac{4\pi t}{U}\right)}\; .
\label{lbound}
\end{equation}

\noindent
On the other hand, in order a nonperturbative nature of CDW potential scattering
would come into power and prowide discrete features seen in Figs. 1-2, the potential
energy cost of a CDW formation with finite amplitude should not be too high. Namely,
the argument of the Bessel functions in Eqs. (\ref{chi3}), (\ref{chibes}) should be
allowed to be of order unity, i.e. $U\rho(\delta_{c})\geq t\delta_{c}$. Substituting
then the resulting value of $\rho$: $\rho(\delta_{c})\geq t\delta_{c}/U$, into the
 potential energy term in Eq. (\ref{ometil}) and requiring that it would be less
than temperature $T\sim T_{SDW}$ one arrives at the upper bound for $x_{d}$:

\begin{equation}
x_{d}\ll \displaystyle\sqrt{\frac{U}{t}}\exp{\left(-\frac{2\pi t}{U}\right)}\; .
\label{ubound}
\end{equation}

\noindent
Combining Eqs. (\ref{lbound}) and (\ref{ubound}) we find an interval of `allowed'
values of the doping concentration $x_{d}$, where the `matching pairs' effect could
be expected:

\begin{equation}
\displaystyle \exp{\left(-\frac{4\pi t}{U}\right)}\ll x_{d} \ll\displaystyle
\sqrt{\frac{U}{t}}\exp{\left(-\frac{2\pi t}{U}\right)}\; .
\label{inter}
\end{equation}

\noindent
As the $x_{d}$ interval above is defined purely via $U/t$ ratio, i.e. coupling
strength parameter, we see that in fact Eq. (\ref{inter}) defines a region in the
 phase diagram in the
coordinates $x_{d}$, $U/t$, inside which the effect predicted in this paper should take
place.

According to Figs. 1-2 the most probable CDW fluctuation at the temperatures T close
to T$_{SDW}$ has wave-vector $\delta_{c}=2\delta$. Also, according to Fig. 3, the
lowest order SDW fluctuation coupled to this CDW has wave-vector $\delta_{s}=\delta$.
For the system of a 2-dimensional array of weakly coupled chains (parallel to the
$x$-axis) this (matching) pair of wave-vectors: $1/2 \pm \varepsilon$  and
 $\pm 2\varepsilon$,
translates into a 2-D form: the SDW wave vector is now $(1/2\pm\varepsilon,1/2)$ ,
 and the
CDW wave vector is $(\pm 2\varepsilon,0)$, which coincides with the stripe-theory
prediction in a moderately strong coupling limit, $U/4t\geq 1$,
for the square
 lattice,
\cite{6}.
This coincidence is nevertheless quite limited, in the sense that here we have found
the most probable configuration of the coupled spin-charge {\it{fluctuations}} above
T$_{SDW}$ (see also \cite{0}), while
theory \cite{6} describes the {\it{ground state}} spin-charge configuration. On the
other hand, when translated into 2-D form, our present result is in
accord with the inelastic neutron scattering data in underdoped cuprates \cite{1} at
the temperatures T greater than superconducting transition temperature T$_{c}$.
Namely, the most intensive magnetic neutron scattering is measured at 2-D wave-vector
$(1/2\pm\varepsilon,\;1/2)$, while  scattering of neutrons by the crystal lattice
 potential
(which indirectly reflects the presense of a CDW) has occured at
 $(\pm 2\varepsilon,\;0)$.
Nevertheless, as in the moderately strong coupling stripe theory \cite{6}, our
prediction  for the doping dependence of $\varepsilon$: $\varepsilon=x_{d}/2$, differs by a
factor of two relative to the experimental result in cuprates : $\varepsilon=x_{d}$.

\section{Stripe phase ordering as a quantum interference phenomenon.}

The main content of this section is based on the work \cite{muk}.
It is well established, that in the Galilean invariant 1D system one finds on the
mean-field level a single mode
Peierls wave \cite{5} with wave vector $2k_{F}$ at arbitrary electron density
($k_{F}$ is the Fermi momentum). It is generally assumed that the same holds true on a
1D lattice away from points of lattice commensuration since then an umklapp
scattering of electrons by inverse lattice wave-vector $2\pi/a$ is ineffective
\cite{6b,7}.
As a result, Peierls state with incommensurate periodic spin density
structure (SDW) is predicted in repulsive case both on the mean-field as on the single
loop renormalization group (``parquet'') level. In fact this assertion is incorrect.
Here we demonstrate that in 1D even in a weak coupling limit, $U/t\rightarrow 0$, the
spin-charge modes coupling is relevant at small doping $x_{d}$, i.e. close to
half-filling of electron band ($1-x_{d}$ electron per lattice site). The mechanism is
genuinely novel \cite{muk} and is related to interference phenomenon involving multiple
scatterings
of electrons by inverse lattice wave-vector and by wave-vector of self-consistently
generated charge density mode (CDW). Essential role of the umklapp scattering makes
this phenomenon strongly doping-dependent as distinguished from welknown phenomenon
of related $2k_{F}$-SDW and $4k_{F}$-CDW fluctuations in a Luttinger liquid \cite{6}.

Recently the Landau free-energy functional was phenomenologicallly introduced
\cite{4}, which contains a leading order mode coupling term between the fundamental
Fourier components of the spin ($\vec{S_{q}}$) and charge ($\rho_{k}$) order
parameters :

\begin{equation}
{\cal{F}}=\frac{1}{2}r_{\rho}|\rho_{k}|^{2}+\frac{1}{2}r_{s}|\vec{S}_{q}|^{2}+
\lambda_{1}
[(\vec{S}_{q}\cdot\vec{S}_{q})\rho_{k}^{*}+\text{c.c.}]+\quad \text{quartic terms}
\label{zek}
\end{equation}

\noindent
with condition imposed: $\vec{k}=2\vec{q}$.
Here the third term is relevant signature of the higher harmonics coupling as well.
Mean-field calculations perfomed for 2D t-U Hubbard models \cite{2,3} show that
spin-charge modes coupling is important at intermediate to strong couplings range
of U/t. What happens in 1D case, especially at weak coupling? In fact, in this case
the mode couplings are important at arbitrary small U/t (at the least if doping
is small but enough to drive the system away from commensuration), and to make this
evident a generalization of the usual mean-field strategy is needed, which will be
discussed below.

It is wellknown \cite{9}, that formation of SDW with wave vector $Q_-=2k_{F}$, which
connects the opposite Fermi points, see Fig. 5, may be regarded as Bose-condensation of
electron-hole pairs, $c_{k,\sigma}c^{\dagger}_{k+Q_-,\sigma}|O\rangle$, with the
binding energy determined by the $Q_-$-scattering amplitude (here $|O\rangle$ denotes
unperturbed vacuum state of the Fermi system). In the presence of
the lattice potential, the scattering of electrons is split into  $Q_-$- and
$Q_+ =2\pi/a-2k_{F}$-scattering processes, the latter involves Umklapp process. In the
commensurate case, i.e. at half-filling, a
constructive interference between the two processes takes place, see Fig. 5a. In the
incommensurate case, i.e. at finite doping: $x_d=2\pi/a-4k_F$, the Umklapp based
$Q_+$-scattering channel brings electron away from the Fermi point region, thus making
minor contribution to the electron-hole coupling, see Fig. 5b. Electron
scattering by a {\it{self-consistently generated}} CDW with small wave-vector
$Q_{CDW}=Q_+-Q_-=x_d$ may (partially) restore constructive interference between
$Q_-$- and $Q_+$-scattering processes at some ``proper'' phase shift $\phi$
between SDW and CDW. Hence, free
energy of thus formed ``stripe phase'', i.e. SDW and CDW with the fixed phase
shift \cite{1,2,3,4}
might be lower than that of a single incommensurate SDW state \cite{muk}. Finally,
electron scattering due to a $4k_F$-CDW potential, unlike due to the long-wavelength
$Q_+-Q_-$-CDW potential mentioned above, would
interfere with the SDW-induced $2k_{F}$-scattering process at any doping $x_d$, also
in the absence of the lattice potential, i.e. in the ``empty lattice'' case, see
Fig. 5c.

In the presence of the $2k_F$-SDW and $Q_+-Q_-$-CDW condensates, $m(x)$ and
$\rho(x)$, the
single-particle eigenstates of the t-U Hubbard Hamiltonian
in the Hartree-Fock approximation can be determined from the Bogoliubov-de Gennes
equations derived in \cite{3}:

\begin{eqnarray}
\mp i2t\frac{\partial u_{\pm}}{\partial x} +
\frac{U}{2}\rho(x)u_{\pm}-\frac{U}{2}\tilde{m}(x)u_{\mp}= Eu_{\pm}
\label{Gennes}
\end{eqnarray}

\noindent
Here left- and right-movers representation (for momenta close to undoped
Fermi ``surface''points $\pm\pi/2a$) is used for the quasi-particle wave
function: $\psi_{\sigma}(x)=u_{+}(x)\exp{(ix/4)}+\sigma u_{-}(x)\exp{(-ix/4)}$;
and the wave-vectors are expressed in units of $2\pi/a$.
The spin density is:
$m(x_{i})/2=(m_{o}/2)\cos{(Q_{-}x_{i}-\phi)}=(m_{o}/2)\cos{(\varepsilon x_{i}+\phi)}
\cos{(x_{i}/2)}$,  where $2\varepsilon\equiv Q_{+}-Q_{-}=2\pi/a-4k_{F}=x_{d}$. Hence,
SDW modulation is: $\tilde{m}(x)\equiv(m_{o}/2)\cos{(\varepsilon x+\phi)}$, and
charge density varies as:
$\rho(x)=\rho_{o}\cos{2\varepsilon x}$. Thus, only slowly varying functions
$u_{\pm}(x)$, $\tilde{m}(x)$ and $\rho(x)$ are involved in Eq.(\ref{Gennes}).

Now, exploit a mathematical trick used in \cite{lebed} for the problem of
field-induced spin density wave (FISDW) in quasi-1D (TMTSF)$_{2}$X compounds.
Unlike in \cite{lebed}, here we consider CDW potential instead of magnetic
field induced vector-potential encountered in FISDW case.
Namely, we express wave functions $u_{\pm}(x)$ in the Bloch-wave basis of
the periodic CDW potential $U\rho(x)\equiv U\rho_{o}\cos(2\varepsilon x)$:

\begin{eqnarray}
&&u_{\pm}(x)\equiv c^{k}_{\pm}\exp{\left\{ikx\mp
iU/(2v)\int^{x}\rho_{o}\cos(2\varepsilon x')dx'\right\}}\nonumber\\
&&=c^{k}_{\pm}\exp{(ikx)}\exp{(\mp iU\rho_{o}\sin{(2\varepsilon x)}/(4v\varepsilon))}
\nonumber\\
&&=c^{k}_{\pm}\exp{(ikx)}\left\{J_{0}(z)+\sum_{n=1}^{\infty}\left(e^{in2\varepsilon x}+
(-1)^{n}e^{-in2\varepsilon x}\right))J_{n}(\mp z)\right\}
\label{Bessel}
\end{eqnarray}

\noindent
where $v=2t$ is the Fermi velocity of electrons, and $z=U\rho_{o}/(4v\varepsilon)$.
Here expansion of $\exp{\{i\sin{x}\}}$ in Bessel functions $J_{n}$ of integer order $n$
was used. Since $J_{n\geq 1}(z)\propto z^{n}$ when $z\rightarrow 0$, we retain only
zero- and first order Bessel functions in
the last line in Eq. (\ref{Bessel}) when $U\rho_{o}/(2v\varepsilon) < 1$.
After substitution of $u_{\pm}(x)$ from Eq. (\ref{Bessel}) into
Eq. (\ref{Gennes}) one finds an algebraic system of linear homogeneous equations for
the coefficients $c^{k}_{\pm}$. Corresponding determinant equation defines
the single-particle spectrum \cite{rem}:

\begin{eqnarray}
&&E_{k} = \displaystyle -\frac{v\varepsilon}{2}\pm
\sqrt{\left(k+\frac{\varepsilon}{2}\right)^{2}v^{2}+\Delta^{2}}
                                                        \label{spec}
\\
&&\text{where:}\quad
\Delta\equiv\frac{Um_{o}}{4}f\left(\frac{U\rho_{o}}{2v\varepsilon}\right)\quad;
\nonumber\\
&&f^{2}(z)\equiv J_{0}^{2}(z)-2\cos(2\phi)J_{0}(z)J_{1}(z)+J_{1}^{2}(z)\quad .
                                \label{f}
\end{eqnarray}

\noindent
In the ``electron doping'' case the sign in front of $\varepsilon$ in Eq.(\ref{spec})
and of $\cos{2\phi}$ in Eq. (\ref{f}) should be changed.
The physical implication of Eqs. (\ref{spec}), (\ref{f}) is remarkable. Namely,
in the absence of CDW $z=0$; hence, $f(0)\equiv 1$ and
the gap $\Delta$ equals $Um_{o}/4$  (at $z\rightarrow 0$ the Bessel functions behave
as: $J_{0}(z)\approx 1-z^{2}/4$, and $J_{1}(z)\approx z/2+{\cal{O}}(z^{3})$).
But, in the presence of the CDW, $\Delta$ is enhanced when $f(z)>1$.
The latter condition fixes sign of $\cos(2\phi)$ in Eq. (\ref{f}).
The form of Eq.(\ref{f}) manifests
quantum interference between scattering amplitudes of electron in the combined
periodic potentials of $Q_{\pm}$-SDW and matching $(Q_{+}-Q_{-})$-CDW. Indeed,
r.h.s. of Eq. (\ref{f}) for $f^{2}(z)$ could be rewritten as:
$|J_{0}(z)+J_{1}(z)e^{i(\phi+\pi)}|^{2}$. This is nothing but interference
intensity between amplitudes of electron (hole) scattered ``back'' and ``forward''
by vectors $Q_{-}$ and $Q_{+}$ respectively, with a phase-shift $\phi+\pi$.
Due to mismatch between the scattering wave-vectors, $Q_{+}-Q_{-}\neq 0$, the
interference vanishes in the absence of the matching CDW potential, since then:
$J_{1}(\rho_{o}=0)=0$ and $J_{0}(\rho_{o}=0)=1$.
A simple form of solution
(\ref{spec}) is valid in the weak coupling limit, $U\ll t$, not too close to half
filling, i.e. when $x_{d}\gg \Delta/t$.

Free energy of the system (per unit of length), $\Omega$, at finite temperature
$T(\equiv\beta^{-1})$, follows from Eq.(\ref{spec}) and the Hartree-Fock form
\cite{3} of the Hubbard Hamiltonian:

\begin{eqnarray}
\Omega = \displaystyle (U/8)(m_{o}^{2}/2 +\rho_{o}^{2})-(4T/\pi v)
\displaystyle\int^{E_{b}}_{0}
\ln{\left[2\cosh{(\beta E(\xi)/2)}\right]}\, d\xi
\label{energy}
\end{eqnarray}

\noindent
where $E(\xi)=\sqrt{\xi^{2}+\Delta^{2}}$, and $E_{b}(\sim t)$ is the upper cutoff of
the electron energy.  Expansion of $\Omega$,
Eq.(\ref{energy}), in powers of small CDW and SDW amplitudes yields:

\begin{eqnarray}
&&\Omega\approx \displaystyle \frac{U}{8}(m_{o}^{2}/2
+\rho_{o}^{2})-\frac{\Delta^{2}}{\pi v}\left(
\ln{\left(\frac{2E_{b}}{1.76 T}\right)}-\frac{0.053\Delta^{2}}{T^{2}}\right)\nonumber\\
&&\approx \frac{U}{8}\left(\frac{m_{o}^{2}}{2}\displaystyle
\left\{1-\frac{U}{\pi v}\ln{\left(\frac{2E_{b}}{1.76 T}\right)}\right\}+
\rho_{o}^{2}\right)+\lambda_{1}\rho_{o}m_{o}^{2}+\quad \cdots
\label{expand}
\end{eqnarray}

\noindent
where $\lambda_{1}=-\{U^{3}/(8\pi v^{2}x_{d})\}\ln{\left[2E_{b}/(1.76 T)\right]}$, and
$m_{o}$ and $\rho_{o}$ are amplitudes of SDW and CDW harmonics with
wave-vectors $\varepsilon$ and $2\varepsilon$ respectively. We see that Eq. (\ref{expand})
just recovers phenomenological Landau-Ginzburg functional in Eq. (\ref{zek}),
considered in \cite{4}. Thus, our derivation reveals the quantum interference
nature of the SDW-CDW coupling term in Eq. (\ref{zek}), with coupling constant
$\lambda_{1}$ following from the microscopic theory, Eq. (\ref{expand}).
Notice, that
$|\lambda_{1}|$ {\it{increases when doping decreases}}.

Here we merely list the main results, obtained by minimizing free
energy Eq.(\ref{energy}) with respect to SDW and CDW amplitudes $m_{o}$ and $\rho_{o}$.

i) Coming from the high temperature limit, $\Delta =0$, the SDW-CDW superlattice
condenses with $\cos{(2\phi)}=-1$ or $\cos{(2\phi)}=1$ depending on the sign
of $x_{d}$.
Thus, the nodes of the spin density coincide with the minima (maxima) of the charge
density $\rho(x)$ in the case of the hole (electron) doping, in
accord with the stripe phase topology considered in the strong coupling limit
\cite{2},\cite{3}.

ii) Dimensionless mode coupling strength, $U\rho_{o}/(2\pi tx_{d})$, in the effective
theory Eqs. (\ref{energy}) and (\ref{expand}), grows up to $\sim 1$ at small $x_{d}$,
below $ x_{o}\sim\sqrt{t/U}\exp{(-2\pi t/U)}$. Formal divergence of $\lambda_{1}$ in
Eq. (\ref{expand}) at $x_{d}\rightarrow 0$ signals that higher order
harmonics have to be considered as stabilizing solitonic lattice (compare \cite{13}).
While transition to solitonic regime is governed by parameter
$U\rho_{o}/(2\pi tx_{d})$, the bare coupling constant,
$U/t\ll 1$, remains small. Simultaneously, transition temperature, T$_{c}$,
monotonically increases from the {\it{lowest}} value
$T_{SDW}= 2(\gamma/\pi)t\exp{(-2\pi t/U)}$ at $|x_{d}|\gg x_{o}$, to the {\it{highest}}
value $T_{m}= 2(\gamma/\pi)t\exp{(-2\pi t/(Uf^{2}_{m}))}$ at small doping,
$|x_{d}|<x_{o}$. Here $\gamma=1.78$, and  the maximum value of the function
$f^{2}(z)$ in Eq.(\ref{f}), $f^{2}_{m}\equiv f^{2}(z_{m})\approx 1.5$, is
reached at $z=z_{m}\approx 0.83$ \cite{14}. The increase of T$_{c}$ is
accompanied by
a substantial increase of the SDW amplitude at zero temperature, see Fig. 6.

\noindent
iii) The character of the phase transition changes at $x_{o}$ from the
{\it{first order}} ($x_{d}<x_{o}$) to the {\it{second order}} ($x_{d}>x_{o}$),
Fig. 7. The jumps of
the CDW and SDW amplitudes at the first order transition temperature are:
$m_{o}^{2}\approx x_{d}z_{m}T_{m}\sqrt{2\pi tU}/(Uf_{m})^{2}$ and
$\rho_{o}\approx 2\pi x_{d}z_{m}t/U$. Hence, in the $x_{d}<x_{o}$ region the mode
coupling strength is: $U\rho_{o}/(2\pi tx_{d})\approx 0.83$ (i.e. not $\ll 1$).
Therefore,
our results in this region, based on the neglect of the
higher order SDW/CDW harmonics, might be considered as qualitative rather than
quantitative.
In the II-nd order phase transition region, $|x_{d}|\gg x_{o}$, the order
parameters close to $T_{c}$ behave as: $\rho_{o}\approx
3\tau T_{SDW}^{2}/(x_{d}tU)$, and $m_{o}\approx 12\sqrt{\tau}T_{SDW}/U$; in
qualitative agreement
with \cite{4} (here $\tau\equiv 1-T/T_{SDW}$).

The phase transitions described in ii), iii) above belong to the ``spin-charge
coupling driven'' and ``spin driven'' kinds, in the terminology introduced in \cite{4}.

\subsection{STRIPE PHASE AND "PARQUET" 1D RENORMALIZATION GROUP TECHNIQUE}

An important issue for the (quasi) 1D systems is the influence of fluctuations. We
study it within a single-loop renormalization group (RG) scheme, so-called
``parquet'' approximation \cite{6}, which we adjust for the case of the two order
parameters (SDW/CDW) coupled already on the mean-field level. Conventionally,
``parquet'' - RG equations describe behavior of the two-electron scattering vertices
$\gamma_{1}(\xi)$, $\gamma_{2}(\xi)$, and $\gamma_{3}(\xi)$, accounting for  back-,
forward- , and umklapp scattering of electrons respectively close to the
Fermi ``surface'' points: $\pm k_{F}$. The RG variable, $\xi$, is the logarithm of the
infrared cutoff of the energy/momentum transfer. It is involved in the
(logarithmically)
diverging corrections to the vertices, which are initially defined in the Born
approximation: $g_{i}\equiv \gamma_{i}(\xi=0)\ll 1$. Within ``parquet'' approach only
corrections of the highest power in $\xi$ are retained in each order of the
perturbation
expansion in each $g_{i}$ and then summed to an infinite order. Transition of
the electron system to a strong coupling regime is signalled by divergences of the
vertices $\gamma_{i}(\xi)$ at some finite value $\xi_{o}$ (where ``parquet''
approximation actually fails). In the case of the Hubbard
Hamiltonian at half filling: $g_{i}=U/(4\pi t)$, $i=1,2,3$ and $\xi_{o}=2\pi t/U$
\cite{6}. Away from half filling the
 umklapp condition for two-electron scattering:
$p_{1}+p_{2}=p_{3}+p_{4}\pm 2\pi/a$, can not be fulfilled when all the quasi-momenta
of electrons (before and after scattering) are close to the Fermi surface. In
conventional ``parquet'' theory \cite{6} a strong coupling transition does not occur in
 repulsive 1D
system away from half filling. The reason is that e.g. in the (hole) doped case
the deficiency of momentum transfer: $2\pi/a-4k_{F}\equiv 2\varepsilon=x_{d}\neq 0$,
provides a ``natural'' momentum cutoff, such that at $\xi>\xi_{d}\equiv\ln{(1/x_{d})}$
the growth of $|\gamma_{i}(\xi)|$ terminates. In order to probe the system
for a stripe phase instability in this case, we modify ``parquet'' treatment by adding
infinitesimal probe vertices $\tilde{\gamma}_{i}(\xi)$, which acquire
``starting'' values $\tilde{\gamma}_{i}(\xi_{d})$ at $\xi=\xi_{d}$. The vertices
$\tilde{\gamma}_{1,2}(\xi)$ describe ``umklapp'' scattering with the wave vector
$2\varepsilon=x_{d}\ll 4k_{F}$, brought by the incommensurate CDW component of the
(anticipated) stripe superlattice;
while $\tilde{\gamma}_{3}(\xi)$ is due to combined, (commensurate) lattice- and the CDW
umklapp: $2\pi/a-2\varepsilon\equiv 4k_{F}$. These vertices,
in combination with the commensurate (bare) umklapp vertex $g_{3}$, restore possibility
of umklapp away from half-filling.
Thus, ``enriched'' RG-``parquet'' equations in the interval $\xi_{d}<\xi<\infty$
become:

\begin{eqnarray}
&&\dot{\gamma}_{3}(\xi)= -2\tilde{\gamma}_{3}(\xi)\tilde{\gamma}_{4}(\xi);\;
\dot{\tilde{\gamma}}_{4}(\xi)=-4 Re(\gamma_{3}(\xi)\tilde{\gamma}_{3}^{*}(\xi))
\nonumber\\
&&\dot{\gamma}_{4}(\xi)=-2\tilde{\gamma}_{3}(\xi)\tilde{\gamma}_{3}^{*}(\xi);\;
\dot{\tilde{\gamma}}_{3}(\xi)=-2\tilde{\gamma}_{3}(\xi)\gamma_{4}(\xi)
                                           \label{parquet}
\end{eqnarray}

\noindent
where $\gamma_{4}(\xi)\equiv\gamma_{1}(\xi)-2\gamma_{2}(\xi)$, and same relation is
valid
between $\tilde{\gamma}_{4}$ and $\tilde{\gamma}_{1,2}$. Diverging solutions
of Eqs.(\ref{parquet}), see Fig. 8, can be expressed in the analytical form in the interval
$\xi>\xi_{d}$ : $\gamma_{3}(\xi)=
B\cosh{(C\ln{|\tanh{D(\xi-\tilde{\xi}_{o})}|}+\phi_{o})}$; $\tilde{\gamma}_{4}(\xi)=
\pm\sqrt{2(\gamma_{3}^{2}-B^{2})}$; $\gamma_{4}(\xi)=(D/2)
\coth{2D(\xi-\tilde{\xi}_{o})}$; $\tilde{\gamma}_{3}(\xi)=
DC/(\sqrt{2}\sinh{2D(\xi-\tilde{\xi}_{o})})$, where $\tilde{\xi}_{o}\sim \xi_{o}+
0.5(\xi_{o}-\xi_{d})\ln{(2/|\tilde{\gamma}_{i}(\xi_{d})|)}$, and all the constants
are determined from the boundary conditions for $\gamma_{i}$ and $\tilde{\gamma}_{i}$
at $\xi=0$ and $\xi\approx \xi_{d}$ respectively.

Notice, that position of the stripe-phase ordering point, $\tilde{\xi}_{o}$, shifts
logarithmically to infinity when starting values of the probe vertices
$\tilde{\gamma}_{i}(\xi_{d})$ tend to zero. This means that long-range incommensurate
order is absent in 1D, and stripe phase order exists at the transient
energy(time)/length scales brought by the probe vertices.

Summarizing, a quantum interference mechanism of the stripe phase ordering in
repulsive (quasi) 1D electron system is proposed \cite{coul}. Though 1D fluctuations
smear away mean-field predicted long-range stripe order, "parquet" analysis
indicates transient scale stripe-phase correlations away from half-filling.
The mean-field solution is more relevant in the classical spin $S\rightarrow \infty$
limit, and could be stabilized by a coupling to the lattice deformation.




\section{Analytical stripe phase solution for the Hubbard model: $T=0$.}

In this section an exact analytical solution of the Hartree-Fock problem at $T=0$ for a
one-dimensional electron system at and away from half-filling is described, as it was
first found in \cite{mm}.
This solution provides a unique possibility to investigate {\it analytically} the
structure of the periodic {\it spin-charge} solitonic superlattices.
It also demonstrates fundamental importance of the higher order
commensurability effects, which result in special stability points along the axis of
concentrations of the doped holes. Though there is no long-range order in the purely
one-dimensional system due to destructive influence of fluctuations, real cuprates
are three-dimensional, and therefore, the long-range order survives in the ground
state.

It is well known \cite{gru} that at low enough temperatures
quasi-one-dimensional conductors may undergo a Peierls- or spin-Peierls
transition and develop accordingly either a long-range charge- or
spin-order. In the case of a discrete lattice model the commensuration
effects become important \cite{6b}. Exact solutions of related Hartree-Fock
problems led to the picture of a solitonic lattice appearing away from
half-filling of the bare electron band \cite{br,machida}. The nature of a
soliton was determined then by a corresponding order parameter which was
either lattice deformation, or density of electronic spin. Nevertheless,
recently discovered stripe phases in doped antiferromagnets (cuprates and
nickelates) \cite{1,1a,1b}  have attracted attention to the problem of
{\it coupled} spin and charge order pararmeters in the electron systems.
Numerical mean-field calculations \cite{2,2a} suggest a universality of
the spin-charge multi-mode coupling phenomenon in repulsive electronic systems
of different dimensionalities.
On the other hand, those calculations are bound to use small clusters which
often makes them inconclusive.
Hence, we believe,
that one-dimensional mean-field solutions contain universal features
of the  stripe phase, which are stabilized in higher dimensions. We use derived here
single-chain analytical solutions as building blocks for the stripe phase in quasi
two(three)-dimensional system of parallel chains. In this way
we have found that short-range (nearest neighbour) repulsion between doped
holes, in combination with effects of magnetic misfit energy between the
chains, naturally leads to formation of either ``half-filled'' or ``fully filled''
domain walls in the low- or high doping limits respectively.
In both cases these walls separate neighbouring antiphased antiferromagnetic
domains.

The Hubbard Hamiltonian with the hopping integral $t$ and on-site repulsion
$U$ ($>0$) may be written in the form already presented in Eq. (\ref{hubbard}):

\begin{equation}
H=\displaystyle t\sum_{\langle i,j\rangle
\sigma}c^{\dagger}_{i,\sigma}c_{j,\sigma}+ U\displaystyle\sum_{i}\left(\frac{%
1}{4}\hat{n}^{2}_{i}-(\hat{S}^{z}_{i})^{2}\right)  \label{Hubbard}
\end{equation}

\noindent Here an identity: $\hat{n}_{i\uparrow}\hat{n}_{i\downarrow}=\frac{1%
}{4}\hat{n}^{2}_{i}- (\hat{S}^{z}_{i})^{2}$ is used. Operators $\hat{n}%
\equiv\hat{n}_{\uparrow}+\hat{n}_{\downarrow}$ and $\hat{S}^{z}$ are
fermionic density and spin (z-component) operators respectively, and $\sigma$
is spin index. Hamiltonian (\ref{Hubbard}) has convenient form for
Hartree-Fock decoupling in the presence of the two order parameters, i.e.
electron spin- and charge-densities, $\langle\hat{S}^{z}(x)\rangle$ and
$\langle \hat{n}(x) \rangle$ respectively.
Single-particle eigenstates and eigenvalues of the Hamiltonian Eq.(\ref
{Hubbard}) in the Hartree-Fock approximation can be determined from the
Bogoliubov-de Gennes equations derived in \cite{2a}  (see also  a review
\cite{rev-br}):

\begin{equation}
 \left(-i\displaystyle\frac{d}{dx}\hat\sigma_z +\alpha \rho (x) -
\alpha m(x)\hat\sigma_x\right ){\bf{\Psi_\sigma}}=\varepsilon {\bf{\Psi_\sigma}}
\label{h0}
\end{equation}

\noindent where $\hat\sigma_{z,x}$ are the Pauli matrices, $\varepsilon = E/2t$,
$\alpha = U/4t $; the Plank constant is taken as
unity, and the length is measured in the units of the lattice (chain) period $a$. In
these units momentum and wavevector are dimensionless, and velocity and energy
posses one and the same dimensionality.
The vector ${\bf{\Psi_\sigma}}^{T}\equiv (\Psi_{\sigma +},\Psi_{\sigma -})$ is defined
in terms of the right- left-movers $\Psi_{\sigma\pm} $, which constitute the wave
function:

\begin{equation}
\Psi_{\sigma }(x) =\Psi_{\sigma +} (x)e^{ i\pi x/2} + \sigma \Psi_{\sigma -} (x)
e^{-i\pi x/2 },
\label{psi}
\end{equation}
where $\sigma =\pm 1$ for a spin $\uparrow$ and $\downarrow$ respectively. The
Fermi-momentum is $p_F= \pi\bar\rho/2$, where
in the case of half-filling the
average number of electrons per site equals $\bar\rho = 1$, i.e. $p_F=\pi/2$.
The slowly varying real functions $m(x)$ and $\rho (x)$ are defined as:
\begin{equation}
 \langle \hat{n}(x) \rangle =   \rho (x), \qquad
 \langle \hat{S}^{z}(x)\rangle = m(x)\cos(\pi x).
\label{def}
\end{equation}
Note that  spin and  charge densities have the same coupling constant $\alpha$.
This is not a necessary constraint, our results remain valid for a more general case
$\alpha \geq \beta $, where $\alpha$ and $\beta$ are spin and charge coupling
constant, respectively.

For a discrete lattice this gives: $\langle \hat{S}^{z}(i)\rangle = (-1)^i m(i)$.
The total energy is equal to \cite{2a}
\begin{equation}
\frac{W}{2t} = \sum_{\varepsilon< \mu} \varepsilon + \int dx \frac{\alpha}{2}
(m^{2}(x) -\rho^2 (x)),  \label{w}
\end{equation}
where $\mu$ is a chemical potential.
Next, we introduce $\bar \rho $ and $\tilde \rho $ as $\rho (x) = \bar \rho +
\tilde \rho (x)$ and $\int\tilde\rho (x) dx = 0$, and pass to a new basis $\Psi_1 $,
$\Psi_2$, according to \cite{muk} (in what follows we drop spin index $\sigma$) :

\begin{equation}
\Psi_{\pm }(x) = \exp(\mp i\alpha \int^{x} \tilde \rho \, dx' ) \Psi_{1,2}(x),
\label{12}
\end{equation}

\noindent
Using this basis we can rewrite Eq.(\ref{h0}) in the form of a single complex order
parameter

\begin{equation}
 \left(-i\displaystyle\frac{d}{dx}\hat\sigma_z +\Delta (x)\hat\sigma_{+}+
\Delta^{*} (x)\hat\sigma_{-}\right ){\bf{\Psi}}=(\varepsilon-\alpha\bar\rho){\bf{\Psi}}
\label{h}
\end{equation}

\noindent
where: $\Delta (x) = -\alpha m(x) \exp(2i\alpha \int\tilde \rho (x) dx )$,
$2\hat\sigma_{\pm}=\hat\sigma_x +\pm \hat\sigma_y$ and
${\bf{\Psi}}^{T}\equiv (\Psi_{1},\Psi_{2})$.

\noindent
It is easy to find from the finite band potential theory \cite{dubrovin} all
formal single-soliton solutions of the eigenvalue problem (\ref{h}):

\begin{equation}
\Delta (x) = \varepsilon_0 -ik_0 \tanh k_0 x  ;\;\; k_0 = \sqrt{\Delta_0^2 -
\varepsilon_0^2 }\, .
\label{dsol}
\end{equation}

\noindent
Here $\varepsilon_0$ is the energy of the localized level, counted from the chemical
potential, and $2\Delta_0 $ is the gap in the energy spectrum.
Notice, that we consider $\Delta_0$ and $\varepsilon_0$ (or equivalent pair of
variables) as {\it two} independent variational parameters. The reason is that,
according to Eqs. (\ref{h0}) and (\ref{h}), there are {\it two
independent mean fields} ``hidden'' in $\Delta(x)$. Each of them should obey
a self-consistency equation. First, we derive equation for $m(x)$, using
Eq. (\ref{psi}):

\begin{equation}
m(x)=(1/2)\sum_{\sigma, E< \mu}(\Psi_{\sigma +}(x)\Psi_{\sigma -}^{*}(x)+c.c.)
\label{spin}
\end{equation}

\noindent
which can also be rewritten in the $\Psi_{1,2}(x)$ representation.
Definition given in Eq. (\ref{12}) leads to a self-consistency equation for the
variable part of the charge mean-field
$\tilde\rho(x)\equiv \langle \hat{n}(x)\rangle-\bar\rho$ :

\begin{equation}
\tilde \rho (x) = \sum_{E< \mu} (\Psi_1^+(x)\Psi_1(x) +
\Psi_2^+(x)\Psi_2(x))- \bar \rho   \label{rho}
\end{equation}

\noindent
The wave functions of the continuum spectrum are as follows:

\begin{equation}
\Psi_{1,2} =\frac{\pm \varepsilon \mp \varepsilon_0 +k +
ik_o \tanh k_0 x }{2\sqrt{L}\sqrt{\varepsilon (\varepsilon-\varepsilon_0) -k_0 /L}}
e^{ikx},   \label{wv}
\end{equation}

\noindent
where $\varepsilon^2 = \Delta_0^2 + k^2$ , and the upper (lower) sign on the r.h.s.
corresponds to the index ``1''(``2''); $L$
is the length of the system (in units of the lattice spacing).

\noindent
Simultaneously, the wave functions of the localized state with energy $\varepsilon_0$
become:

\begin{equation}
\Psi_1(x) = \Psi_2(x) =\frac{\sqrt{k_0}}{2\cosh k_0 x }.  \label{lwf}
\end{equation}

\noindent
For correct summation over the energy levels in all the equations above, periodic
boundary conditions are imposed on the wave functions $\Psi_{\pm}$ of the continuum
spectrum: $\Psi_{\pm}(x+L) =\Psi_{\pm}(x)$. Then, quantization condition follows:
$kL + \arctan({k_0}/{k}) =2\pi n$, where n is an integer. Resulting self-consisting
charge and spin components of the single-soliton state are obtained:

\begin{eqnarray}
&&\displaystyle\tilde \rho (x)= \frac{k_0}{2\cosh^2 k_0 x }\Gamma,\quad
\Gamma\equiv \nu - 1 + \frac{2}{\pi}\arcsin\frac{\varepsilon_0}{\Delta_0},
\label{trho}\\
&&\displaystyle m(x) = -  \frac{\varepsilon_0 \pm ik_0 \tanh k_0 x }{\alpha }
\exp{\{\pm 2i\alpha\phi(x) +i\chi\} } ,  \label{m}
\end{eqnarray}

\noindent
where $\phi(x)=\int^{x} \tilde \rho \, dx'$, $\nu $ is the filling factor of the
localized level ($\nu = 0, 1, 2 $); and
$\chi $ is an arbitrary phase.
As is evident from Eq. (\ref{def}), at half filling, i.e. in the commensurate case
$\bar\rho = 1$, the order parameter $m(x)$ must be real.  In order to fulfil this
condition in Eq. (\ref{m}) (to the lowest order in $\alpha$ for the case $\nu =0, 2$,
and precisely, for $\nu = 1$) one chooses: $\varepsilon_0=0$. Eqs. (\ref{trho}),
(\ref{m}) describe structures of the topological spin-charge solitons (kinks), which
are
either {\it spinless} with charge $\pm e$ (single electron charge) at $\nu = 0,2$, or
{\it chargless} with spin $1/2$ in the case $\nu=1$. In all the cases there are two
antiphased antiferromagnetic domains in the system.  The  $\nu=1$ soliton is the
stationary excited state of the undoped system. The same is true in the cases $\nu=0$
($\nu =2$), but with one electron removed(added) from(to) the system (i.e. still
zero doping in the thermodynamic limit). Important is
that when doping with holes or electrons becomes finite, already the {\it ground state}
of the system  possesses periodic ``chain'' of the alternating spinless solitons
and antisolitons of the kind $\nu =0$ or $\nu =2$ (for hole- or electron doping
respectively), which are described above. This
solitonic superlattice, as we
assume, is a one-dimensional analogue of the stripe phase observed in lightly doped
cuprates and nickelates \cite{1,1a,1b}. Existence of the superlattice is
related to appearance of the central band around $\varepsilon=0$, filled with either
spinless ``holes'' ( $\nu =0$) or spinless ``electrons'' ( $\nu =2$).

In order to
find a structure of the ground state at finite doping
$n_{h}\equiv\vert 1-\bar\rho\vert\neq 0$ we calculate the total
energy Eq. (\ref{w}) of the electron system using solutions (\ref{trho}) and (\ref{m}).
We obtain for the potential energy, i.e. the second term in Eq. (\ref{w}):
\begin{equation}
\frac{W_{pot}}{2t} =\frac{%
L\Delta_0^2}{2\alpha } -\frac{k_0}{\alpha} -\frac{\alpha }{6}\Gamma^{2}
k_0-\frac{\alpha}{2}L\bar\rho^2.
\label{wpot}
\end{equation}
\noindent where $\Gamma$ is defined in Eq. (\ref{trho}). The other part of the total
energy  in Eq. (\ref{w}) reads:
\begin{equation}
\frac{W_{el}}{2t} =-\displaystyle\frac{L}{\pi}(p_F\varepsilon_F
 +\Delta_0^2 \ln \frac{\varepsilon_F+p_F}{\Delta_0} ) +w,  \label{wel}
\end{equation}
where $  \varepsilon^2_F = p^2_F + \Delta^2_0$, and
the term $w$ is of the order $L^0$.

Minimization of the total energy with respect to $\Delta_0 $ in the order $%
\propto L$ gives the usual result \cite{br} : $\Delta_0 =
2\varepsilon_F \exp(-{1}/{\lambda})$, where $\lambda ={2\alpha}/{\pi}$.
Since our approximation makes sense when $\Delta_0 \ll \varepsilon_F$, we
conclude that parameter $\alpha = U/4t$ should be much less than 1 (weak coupling
limit). In this limit we see that part in the potential energy $\int\alpha
\tilde\rho^2 /2$ is of the order $\alpha $ and is much less than other
terms, which are of the order $\alpha^{-1}$ or $\alpha^0 $.

The energy of the kink
 state, Eq. (\ref{w}), can be expressed similar to the work \cite{br}:
\begin{equation}
\frac{W}{2t} = \Delta_0 [\gamma\cos \theta + \frac{2%
}{\pi} \sin \theta ] -\frac{\alpha}{6} \Delta_0\gamma^{2}\sin \theta ,  \label{wex}
\end{equation}
where $\gamma \equiv \nu -2\theta/\pi$, and $\cos \theta
=\varepsilon_0/\Delta_0$, $\sin \theta = k_0/\Delta_0 $.
As we have already seen, $\varepsilon_0 = 0$ for the half-filled case, thus leading to
$\theta = \pi /2$.

In the lowest order expansion in $\alpha \ll 1$
the solitonic structure is described in terms of elliptic functions, see Fig. 9:
\begin{eqnarray}
&&m(x) = (\Delta_0/\alpha) \sqrt{q} sn(\Delta_0  x/  \sqrt{q} , q), \label{m1}\\
&& \tilde \rho (x)= {K(r^{\prime})\Delta_0 r}\left[({\alpha m(x)}/
{\Delta_0})^{2}+C\right]/{\pi},  \label{trho1}
\end{eqnarray}
where $C\equiv 1 -{2}(1-{E(r)}/{K(r)}){r^{-2}}$, and
$ sn (\Delta_0  x/  \sqrt{q} , q )$ is the Jakobi elliptic function
with the parameter $0<q<1$ defined by $2K(q)\sqrt{q}/\Delta_0 =
1/\vert \bar\rho - 1\vert $. Here $K(r)$ and $E(r)$ are complete elliptic integrals
 of the first and second kind respectively, and $r = 2\sqrt{q}/(q+1)$,
$r^{\prime}=\sqrt{1-r^2}$. Parameter $q$ varies from $q=1$ at $\bar\rho =1$ where
$ sn(\Delta_0 x) = \tanh (\Delta_0 x)$, to $q \ll 1$ where
$sn (\Delta_0 x/\sqrt{q}, q) \sim \sin(\pi \vert\bar\rho -1 \vert x)$. Simultaneously,
$E(0)/K(0)=1$, and $E(1)/K(1)=0$.

In the limiting case of ``overdoping'': $\vert \bar\rho -1\vert \gg \Delta_0$, in
which case $q\ll 1$ and $K(q)\approx \pi/2$, one has:

\begin{equation}
 m(x) \approx    {\Delta_0^2 }({\pi\alpha \vert \bar\rho -1\vert})^{-1}
 \sin (\pi \vert\bar\rho - 1 \vert x),
\label{m2}
\end{equation}

\begin{equation}
\tilde{\rho}(x) \approx {\Delta_0^4}({\pi^2\vert\bar\rho -1\vert^3})^{-1}
\cos (2\pi\vert\bar\rho -1\vert x),
\label{trho2}
\end{equation}
in qualitative accord with the approximation of the main harmonics coupling used in
\cite{muk}.
Minimization of (\ref{wex}) with respect to $\theta$ in the case $\bar\rho \neq 1$
(it is assumed that $\vert\bar\rho -1\vert \gg \Delta_0 /v_F $, and the ground
state structure is described by Eqs. (\ref{m2}),(\ref{trho2}) leads to the
excited state solutions,
which are the same as for the Peierls model and are independent of $\alpha$ when
$\alpha \ll 1$ . The only nontrivial solution is a solitonic excited state with:
$\nu = 1, \varepsilon_0 = 0, Q = 0, W-W_0 =2\Delta_0/\pi$ ; where $Q$ is the charge
of a soliton. This solitonic excitation corresponds to a gradual phase-shift by
$\pi/2$ of the argument of sine in the ground state solution in Eq. (\ref{m2}).
Thus, we see that the
structure of the ground state of
our model should be similar to the one of  the Peierls model \cite{br}. Hence,
we can conclude that  for the finite hole density $n_h = \vert\bar\rho - 1\vert$
the ground state charged  ($\nu = 0,\ 2$ ) spin-solitons form a periodic
(super)structure with the spin period $l= 2/\vert\bar \rho - 1\vert $ equal twice
the charge period.
It is known that in any exactly integrable model there is no commensurability
effects at the commensurate points, \cite{dz} : $\vert \rho_0 - 1\vert = m/n$,
where $m$, $n$ are relatively prime integers. That is the energy and other
system parameters continuously depend on the filling factor $\bar\rho $. But in
our case the term $-\alpha \rho^2/2$ in the potential energy (\ref{w})
violates exact integrability and high order  "umklapp" processes
 lead to a pinning of the spin density wave
$m(x)$. As a result at any commensurate point we obtain a decrease in the total
energy of the order \cite{dz}: $\delta w \propto -\alpha \exp\{-n\,\, const \}$.

To summarise, we consider briefly the two-dimensional (2D) case using our 1D results
by adding weak interchain interactions (see also \cite{rev-br} for the charge solitons
at half-filling caused by $t_{\perp}$-band effects).
In the lowest order approximation in the
interchain hopping integral $t_{\perp}$ the interaction energy is

\begin{equation}
\delta W=-J\sum_{<i,j>}\int_{0}^{L}dx(\cos (\varphi _{i}-\varphi
_{j})-1)+W_{C},  \label{W}
\end{equation}
where $J\sim t_{\perp }^{2}/\Delta _{0}$, $\varphi _{i}$ is the phase of a
spin-density $m(x)$ on the $i$-th chain, $W_{C}$ is the Coulomb interaction
energy between the charged kinks (solitons). We suppose
for simplicity that the Coulomb interaction decreases rapidly with the distance
and take into account only charged kinks on the neighbouring
sites of the neighbouring chains, $W_{C}=N_{p}Q^{2}/\varepsilon a'$, where
 $N_{p}$ is the number of pairs of the charged kinks, $\varepsilon $ is a
dielectric constant, and $a'$ is an interchain spacing. In the half-filled system, i.e. in
the absence of the charged kinks, the minimum of (\ref{W}) is
achieved when $\varphi _{i}=\varphi _{j}$ for all $i$, $j$, and we have
$\delta W=0$.
There are two possible ways to create a periodic solitonic structure (solitonic
superlattice) in the doped 2D system. In the first case the charged kinks would
reside on every chain, while in the second case only every even (odd) chain would
be doped. Compare the energies of these two different configurations.

In the first case we have $\varphi_i = \varphi_j$ and : $\delta W_1 = W_{C} =
N_h Q^2/2\varepsilon a'$, where $N_h$ is the number of kinks (solitons) and the
 charge $Q$
could be deduced from
Eqs. (\ref{trho}) and (\ref{trho1}). This is a 2D stripe pattern with filled
"domain walls", i.e. having
one spinless charged kink per period $a'$ perpendicular to the chains direction.
In the second case the Coulomb repulsion energy is negligible, but there is
an increment in the total energy due to the first term in (\ref{W}), i.e. due to a
magnetic order misfit: $\varphi_i \neq \varphi_j$. The minimal energy
configuration could be achieved now in two ways. One possible pattern corresponds to
a half-filled "bi-stripe" pattern,  seen experimentally in some doped manganites
\cite{1,1a,1b}.
Namely, the kink - anti-kink pairs of the smallest possible size $\xi$ would form on
each even(odd) chain in order
to keep the phase-shift of $m(x)$ with respect to the odd(even) chains equal to
$2\pi$. The odd(even) chains remain antiferromagnetically ordered.
Then the energy is: $\delta W_2 \sim N_h J\xi$, where $\xi$ is of the order of the
kink width. $\xi$ monotonically increases \cite{ma}
from its value $v_F /\Delta_0 \sim e^{1/\lambda }$ at $\bar\rho =1$, to $v_F
/\Delta \sim e^{2/\lambda }\tan \pi \vert\bar \rho -1\vert/2 $ in the limit
$\vert\bar \rho -1\vert \gg e^{-1/\lambda }$ .
In the case $\delta W_{1}>\delta W_{2}$ we have that the alternating kink
structure is energetically preferable. Notice, that function $\delta W_{1}$
monotonically decreases, while function $\delta W_{2}$ monotonically
increases as the function of $\vert\bar\rho -1\vert$, therefore the sign in the
above inequality may change at some sufficient doping concentration of the holes.

An alternative, half-filled single-stripe pattern may arise when
 $J\xi\gg 2\Delta/\pi$ , i.e. when $t_{\perp}$ is not too small.
In this case the minimum of the energy (\ref{W}) is achieved at $%
\varphi _{i}=\varphi _{j}$. For this to be true for any couple of the chains,
while
$W_{C}\approx 0$ being also fulfilled, the doped holes should again reside, say,
on every even chain in the form of spinless charged solitons $\nu =0$ . But
simultaneously, an equal number of the chargeless solitons
$\nu =1$ (with spin $\pm 1/2$) must
be formed at all the odd chains in order to maintain
the condition $\varphi _{i}=\varphi _{j}$.
This
configuration will be stable if: $W_{s}N_{h}<{Q^{2}}N_{h}/({2\varepsilon a'})$,
where $W_{s}=2\Delta/\pi$ is creation energy of the chargeless kink.
Notice that since the charge $q$ monotonically decreases from 1 to 0 as function
of doping $N_{h}=L\vert 1-\bar\rho\vert$ (see Eqs. (\ref{trho}),(\ref{trho2}))
the above inequality will be not satisfied at high doping densities, and
half-filled to filled stripe transition would be expected in qualitative accord
with experiment \cite{1,1a,1b}.

\section{Analytical stripe phase solution for the Hubbard model: $T>0$.}

Consider thermodynamic properties of the model (\ref{hubbard})
as a function of a chemical potential $\mu$.
The thermodynamic potential has the form
\begin{equation}
\Omega = -T\sum_E \log (1 + {\rm e}^{(\tilde{\mu} - E)/T}) +
 \int dx\, \frac{|\Delta(x)|^2}{2\alpha } -
\int dx \,\frac{\alpha}{2}\rho^2(x),
\label{Om}
\end{equation}
where $\tilde{\mu} = \mu - \alpha \rho$.
An analytical treatment is possible at low temperatures $T \ll \mu,
\Delta_0$,
or near a phase transition where the gap in the spectrum $\Delta \ll T$.

We use the solution
\begin{equation}
\Delta (x) = \Delta_0 \sqrt{k} {\rm sn}(\Delta_0 x/\sqrt{k}, k),\qquad
k = (E_+ - E_-)/(E_+ + E_-),
\label{sn}
\end{equation}
which is an exact one at the limit $\alpha \ll 1$.
The electron spectrum consists of two gaps ($E_- ^2 < E^2 < E_+^2$)
with the dispersion

\begin{equation}
 \frac{dp}{dE} = \frac{E^2 - E_+^2 E(r)/K(r)}{\sqrt{R(E^2)}},
\quad E_{\pm} = \frac{\Delta_0}{2 \sqrt{k}} ({1} \pm k),
\quad r= \frac{2\sqrt{k}}{k+1},
\label{dis}
\end{equation}
where $K(k),\, E(k)$ are the elliptic integrals of the first and
second kind.
Using the explicit form for the wave functions \cite{dubrovin}
we rewrite the last term in (\ref{Om}) as
\begin{equation}
-\frac{\alpha}{2} \bar{\rho}^2 - \frac{\alpha}{2\pi^2}
\left( \int \frac{f(E) dE}{\sqrt{R(E^2)}}\right)^2 [\bar{\Delta^4} -
(\bar{\Delta^2})^2],
\label{ad}
\end{equation}
where $\bar{f}$ is an average value of the periodic function $f(x)$
over a period. After calculations we obtain
\begin{equation}
\bar{\Delta^4} -(\bar{\Delta^2})^2 =
\frac{\Delta_0^4}{k^2} \left[ -\frac{k'^2}{3} +
 \frac{(4-2k^2)E(k)}{3 K(k)}
-\frac{E^2(k)}{K^2(k)}\right]
\label{}
\end{equation}

Consider the case of low temperatures $T \ll \mu, \Delta_0$.
The thermodynamic potential can be expanded in the parameter
 $\bar{\Delta}_0/T$, where $\bar{\Delta}_0$ is the value of the order
parameter at $T= \mu =0$.
 Calculating with the density of state (\ref{dis})
we obtain after minimization over $\Delta_0$ first terms of
expansion of the thermodynamic potential
\begin{equation}
\Omega = \Omega_0 + \frac{2\Delta_0^2}{\pi}\left[-\frac{1}{4} -
\sqrt{\frac{\pi}{2}}
\left(\frac{T}{\Delta_0}\right)^{3/2} (1 - \frac{1}{K(k)})
-\frac{\delta}{K(k)} -\frac{\delta^2}{2(K(k))^2}\right],
\label{exp}
\end{equation}
where $K \approx \log(4/k^{\prime})$ at
$k \to 1, \, k^{\prime} = \sqrt{1-k^2}$, and
 $\delta = ( \tilde{\mu} - 2\Delta_0/\pi)/(2\Delta_0 /\pi)$.
The period of superstructure (the distance between nearest stripes)
 is $l= 4K(k)\sqrt{k}/\Delta_0$. At low chemical potential
$\mu < \mu_1 (T) $
the minimum of the thermodynamic potential is achieved at homogeneous
phase ($\Delta (x) = const, \, k=1$). With increasing of $\mu$
the homogeneous phase becomes unstable,
 the line of phase transition to the stripe state ( $k\neq 1, \, l<\infty$ )
is given by equation
\begin{equation}
\delta + \frac{\pi \alpha}{24}
 = \sqrt{\frac{\pi}{2}}\left(\frac{T_1}{\Delta_0}\right)^{3/2}
\exp\left[\frac{\mu_0 - \Delta_0}{T_1}\right],\qquad
\mu_0 = 2\Delta_0 /\pi + \alpha \rho.
\label{T1}
\end{equation}
The transition is the phase transition of the first order
due to the negative term $\propto 1/K^2$ in the
thermodynamic potential (\ref{exp}).

The boundary line $T_2 (\mu )$
 between the normal and homogeneous spin density wave state
is determined by taking $k=1, \, \Delta_0 \to 0$ in self-consistent equations
$\partial \Omega /\partial k=0,\, \partial \Omega /\partial \Delta_0 =0$.
In the lowest order in $\alpha$ this line is given by  the equation
\begin{equation}
\log\frac{T_0}{T_2} + Re \left[\Psi\left(\frac{1}{2}\right) -
 \Psi \left(\frac{1}{2}+ \frac{i \mu}{2\pi T_2} \right)   \right] =0,
\label{T2}
\end{equation}
where $T_0 = \gamma \bar{\Delta}_0 /\pi$, $\gamma$ is the Euler's constant,
and $\Psi$ is the Euler's digamma-function.
The phase boundary $T_3(\Omega )$
between normal and stripe phase is determinated by
 setting $\Delta_0, \, k \to 0$ at constant period of superstructure
$l = 2\pi \sqrt{k}/\Delta_0$.

\begin{equation}
\log\frac{T_0}{T_3} + Re \left[\Psi\left(\frac{1}{2}\right) -
\frac{1}{2} \Psi \left(\frac{1}{2}
+ \frac{b + \mu}{2\pi T_3}i \right) - \frac{1}{2} \Psi \left(\frac{1}{2}
- \frac{b - \mu}{2\pi T_3}i\right) \right] = 0,
\label{T3}
\end{equation}

where the wave number of superstructure
 $2 b$ is the solution of the equation
\begin{equation}
Im \left[ \Psi^{\prime} \left(\frac{1}{2}
+ \frac{b + \mu}{2\pi T_3}i \right) -  \Psi^{\prime} \left(\frac{1}{2}
- \frac{b - \mu}{2\pi T_3}i\right) \right]= 0
\label{T33}
\end{equation}
At large chemical potential $\mu \gg \bar{\Delta}_0$ we
 find $T_3 \propto 1/\mu$.

Note that our results are similar to obtained
in Refs.\cite{machida,buz,mach} for models of
 one-dimensional superconductors or charge density waves.
As distinct from previous cases, our model is not exactly solvable,
therefore equations (\ref{T2}) - (\ref{T33}) are valid in the limit
 $\alpha \ll 1$.

The qualitative phase diagram is depicted in Fig.10.
The phase transition between stripe and antiferromagnet state is first
order,
other transition lines are second order.
All three lines of phase transitions are intersected at the Lifshitz point
P.

\section{Acknowledgement}
 The work was supported in part by NWO and FOM (Dutch Foundation for
 Fundamental Research) during S. Mukhin's stay in Leiden, and by RFBR
 grant No. 02-02-16354.

Instructive discussions with Jan Zaanen and S.A. Brazovski are highly
appreciated by the authors.

\newpage

\section{Figure captions}

\noindent
Fig.1: Probability W($Z_{c}$) as function of (normalized) CDW wave-vector
$Z_{c}\equiv\delta_{c}/\delta$; solid blue (dashed blue)
lines: thermal smearing included (excluded). Parameters:
 $2t/U=1$, doping concentration: $x_{d}=0.02$.
 Blue solid/dashed lines: $T=1.2T_{SDW}$, red solid/dashed lines: $T=2T_{SDW}$.

\vspace{3mm}

\noindent
Fig.2: Probability W($Z_{c}$) as function of CDW wave-vector
$Z_{c}$ with the thermal length effect included. Parameters: $T=1.2T_{SDW}$,
 $2t/U=2$, $x_{d}=0.001$ - solid (black)line; $T=1.2T_{SDW}$, $2t/U=1$, $x_{d}=0.02$ -
dotted (red) line.

\vspace{3mm}

\noindent
Fig.3: Manifold of points $\{Z_{c},Z_{s}\}$, which give non-zero
 contribution into $\chi_{\delta_{s}}(\rho(\delta_{c}),\delta_{c})$
in Eq. (\ref{chibes}).
The dashed vertical grid is guide for the eye. See text for details.

\vspace{3mm}

\noindent
Fig.4:  Same manifold as  in Fig.3 but rarefied along $Z_{c}$ axis
for better visibility. Solid circles indicate points with the number of non-zero
terms in Eq. (\ref{chibes}) equal to 2; open circles and squares indicate points with
this number equal to 1. The vertical grid is guide for the eye.  The "columns" of
circles distinguish $Z_{c}$ - values, which are members of the matching pairs. See text
for details.

\vspace{3mm}

\noindent
Fig. 5. Different electron scattering processes in the a) commensurate, b)
incommensurate, and c) ``empty lattice'' cases. Lattice spacing is: $a=1$, and
$\varepsilon = x_d/2$.

\vspace{3mm}

\noindent
Fig. 6. a): SDW ($m_{o}$) and CDW ($\rho_{o}$) amplitudes
as functions of doping $x_{d}$ at zero temperature, calculated using
Eq.(\ref{energy}) with $4t/U=3.2$; b): normalized stripe phase transition
temperature T$_{c}$ as function of doping. Regions of the I-st and II-nd order
phase transition are separated by vertical dashed line.

\vspace{3mm}

\noindent
Fig. 7. Calculated temperature dependences of the stripe phase order parameters,
$m_{o}$ (curves labeled with ``$m$'') and $\rho_{o}$ (curves not labeled) for the
different doping concentrations $x$, with $4t/U=3.2$. Each pair of lines of the
same type shows $m_{o}$ and $\rho_{o}$ for each particular
value of doping $x$.

\vspace{3mm}

\noindent
Fig. 8. Solid lines: numerical solutions of the Eqs. (\ref{parquet}) for the vertices
$\gamma_{3,4}$ (curves 3 and 4 respectively) and $\tilde{\gamma}_{3,4}$ (curves 3'and 4'
respectively), for two different values of $\xi_d$: $\xi_1<\xi_2$. Dashed lines: behavior
at half-filling. Initial values: $g_{1,2,3}=-g_4=0.1$; $\tilde{\gamma}_{3}(\xi_i)=
-\tilde{\gamma}_{4}(\xi_i)=0.01$.

\vspace{3mm}

\noindent
Fig.9. Solitonic spin-charge coupled superstructures for $\alpha\equiv U/(4t)=0.3$.
Only the first harmonics survive at higher doping $n_{h}\equiv\vert 1-\bar\rho\vert$,
in accord with \cite{muk}.

\vspace{3mm}

\noindent
Fig.10. The qualitative phase diagram: the phase transition between stripe and
antiferromagnet state is first order, other transition lines are second order.
All three lines of phase transitions are intersected at the Lifshitz point P.

\newpage

\begin{figure}
 \vbox to 4.0cm {\vss\hbox to -5.0cm
 {\hss\
       {\includegraphics{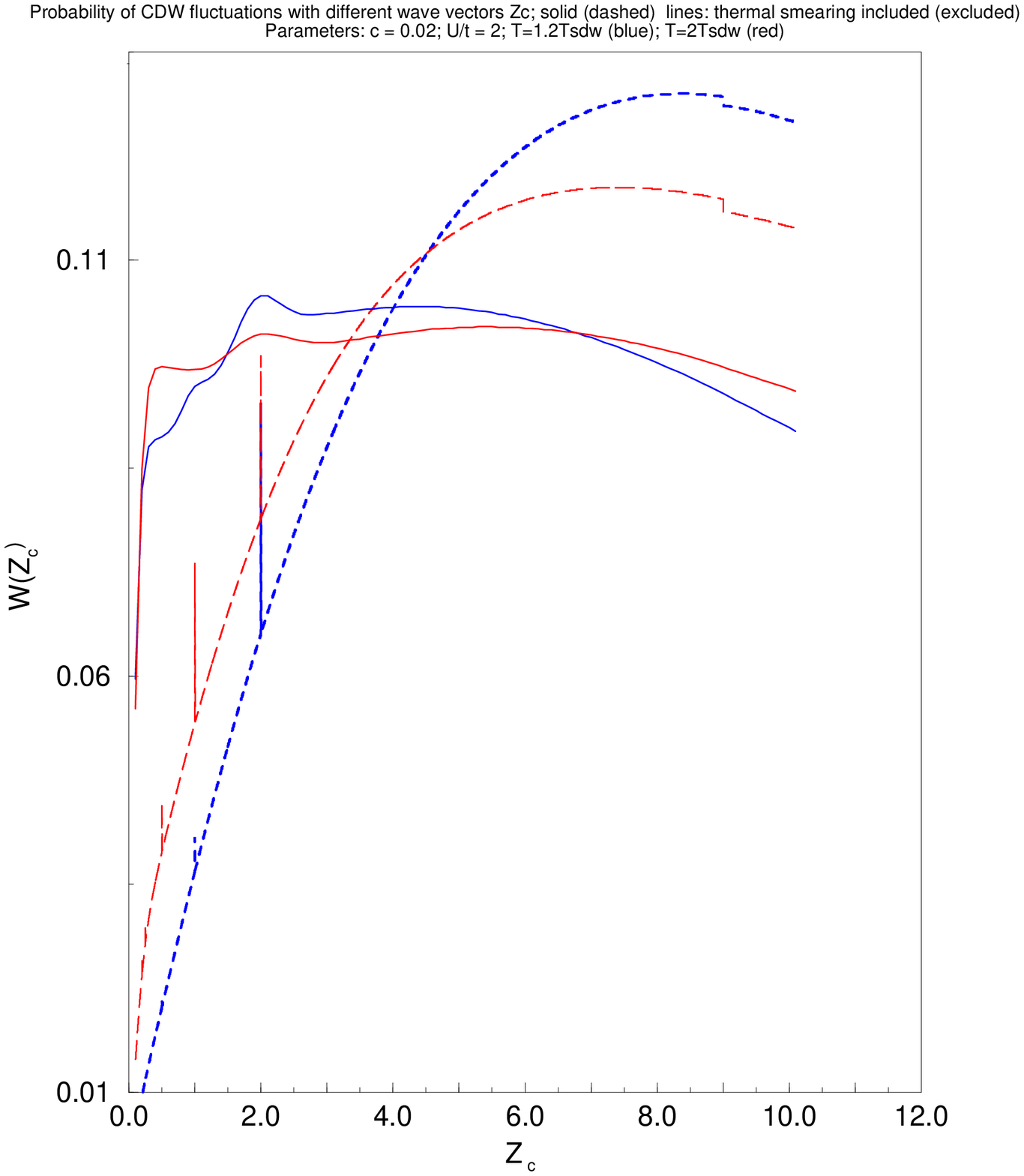}
       }
  \hss}
 }
\vspace{14cm}

\caption{}
\label{fig1}
\end{figure}

\newpage

\begin{figure}
 \vbox to 4.0cm {\vss\hbox to -5.0cm
 {\hss\
       {\includegraphics{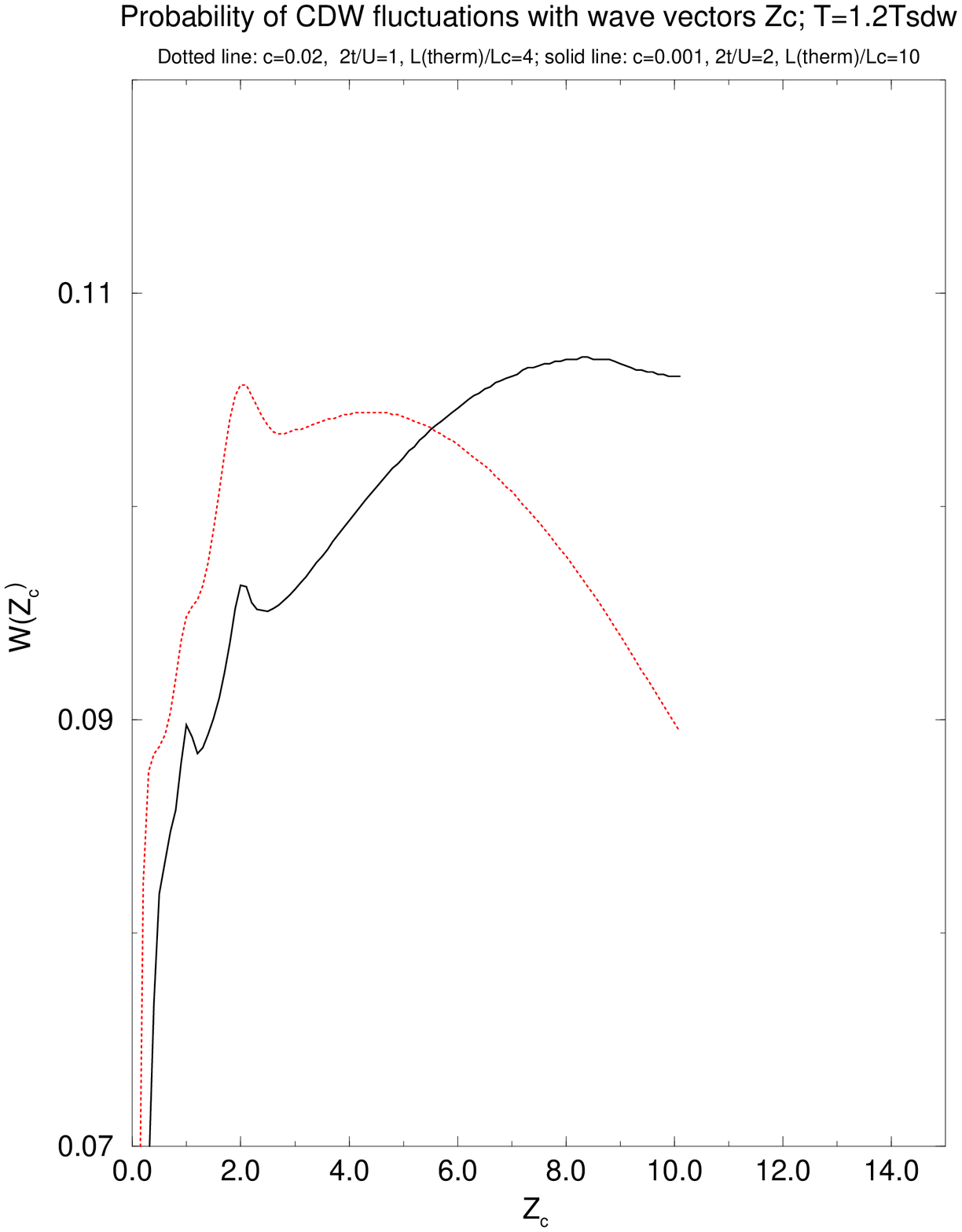}
       }
  \hss}
 }
\vspace{14cm}

\caption{}

\label{fig2}
\end{figure}

\newpage

\begin{figure}
 \vbox to 4.0cm {\vss\hbox to -5.0cm
 {\hss\
       {\includegraphics{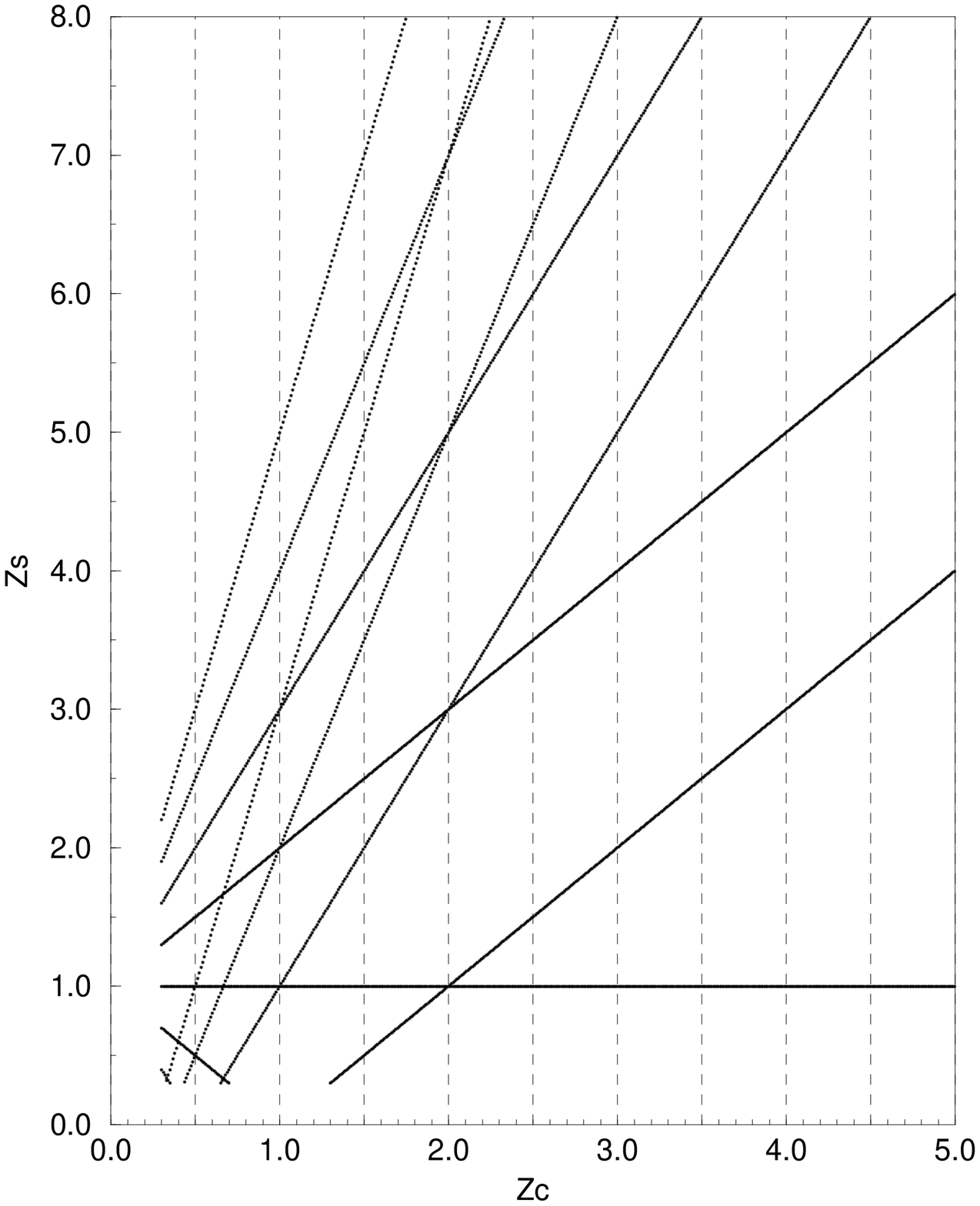}
       }
  \hss}
 }
\vspace{14cm}

\caption{}

\label{fig3}
\end{figure}

\newpage

\begin{figure}
 \vbox to 4.0cm {\vss\hbox to -5.0cm
 {\hss\
       {\includegraphics{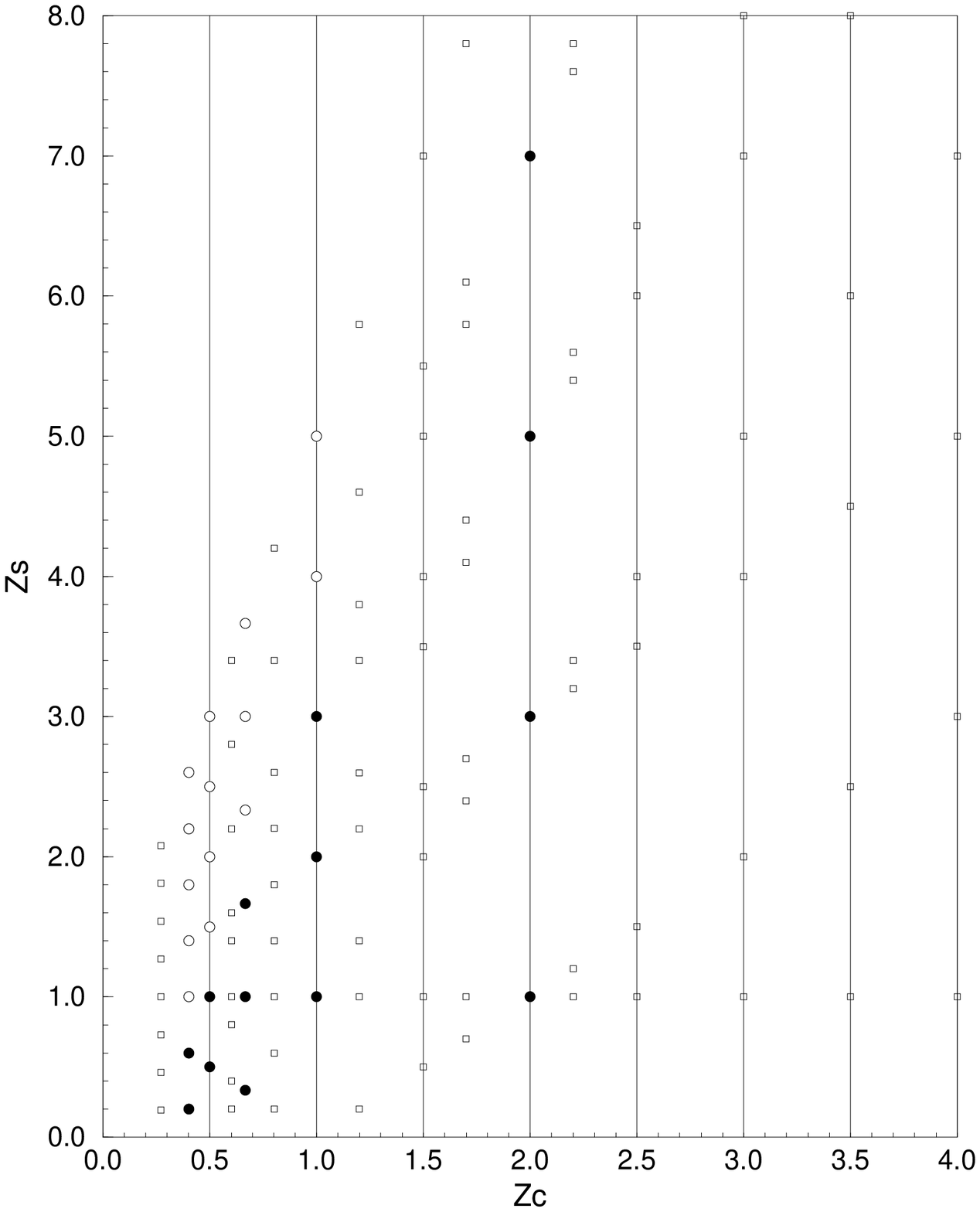}
       }
  \hss}
 }
\vspace{14cm}

\caption{}

\label{fig4}
\end{figure}

\newpage

\begin{figure}
 \vbox to 4.0cm {\vss\hbox to -5.0cm
 {\hss\
       {\includegraphics{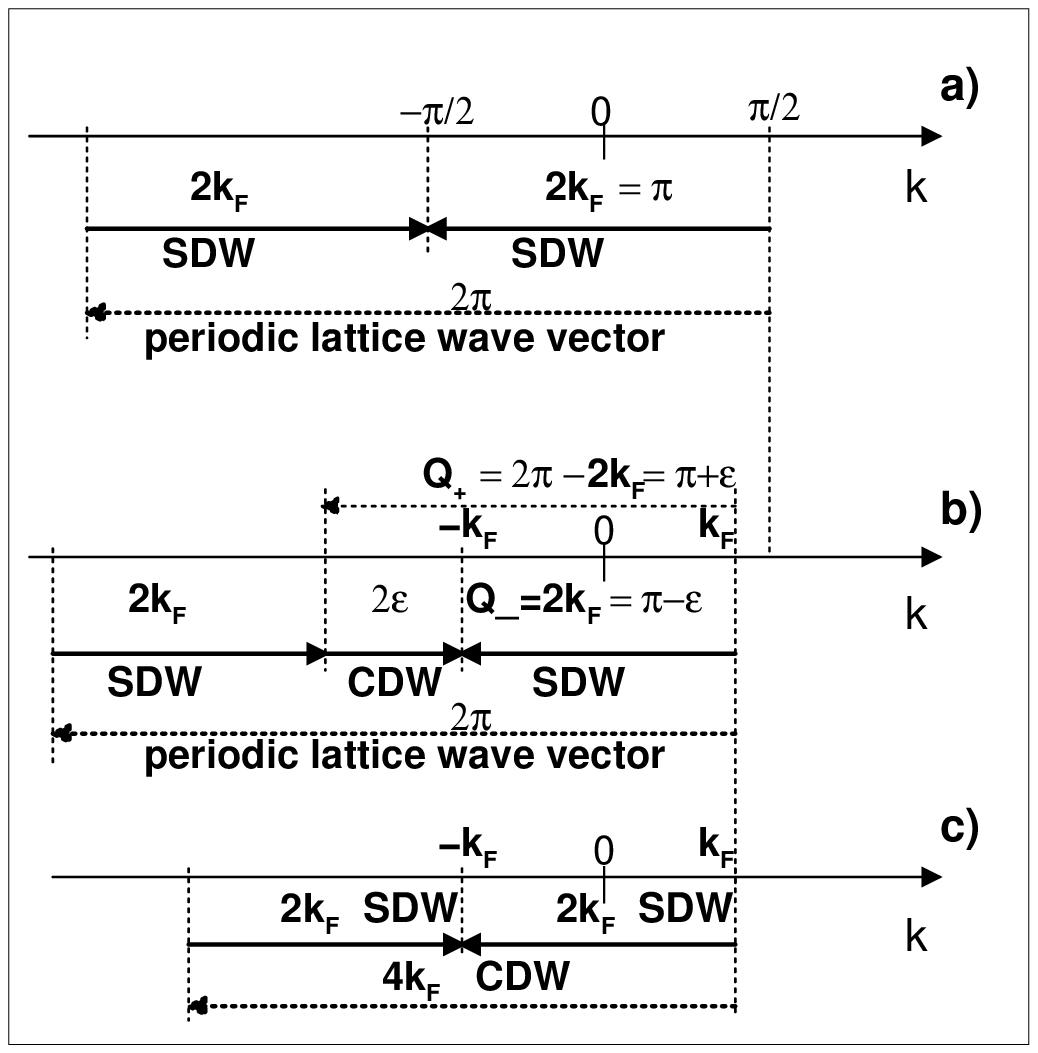}
       }
  \hss}
 }
\vspace{14cm}

\caption{}

\label{fig5}
\end{figure}

\newpage

\begin{figure}
 \vbox to 4.0cm {\vss\hbox to -5.0cm
 {\hss\
       {\includegraphics{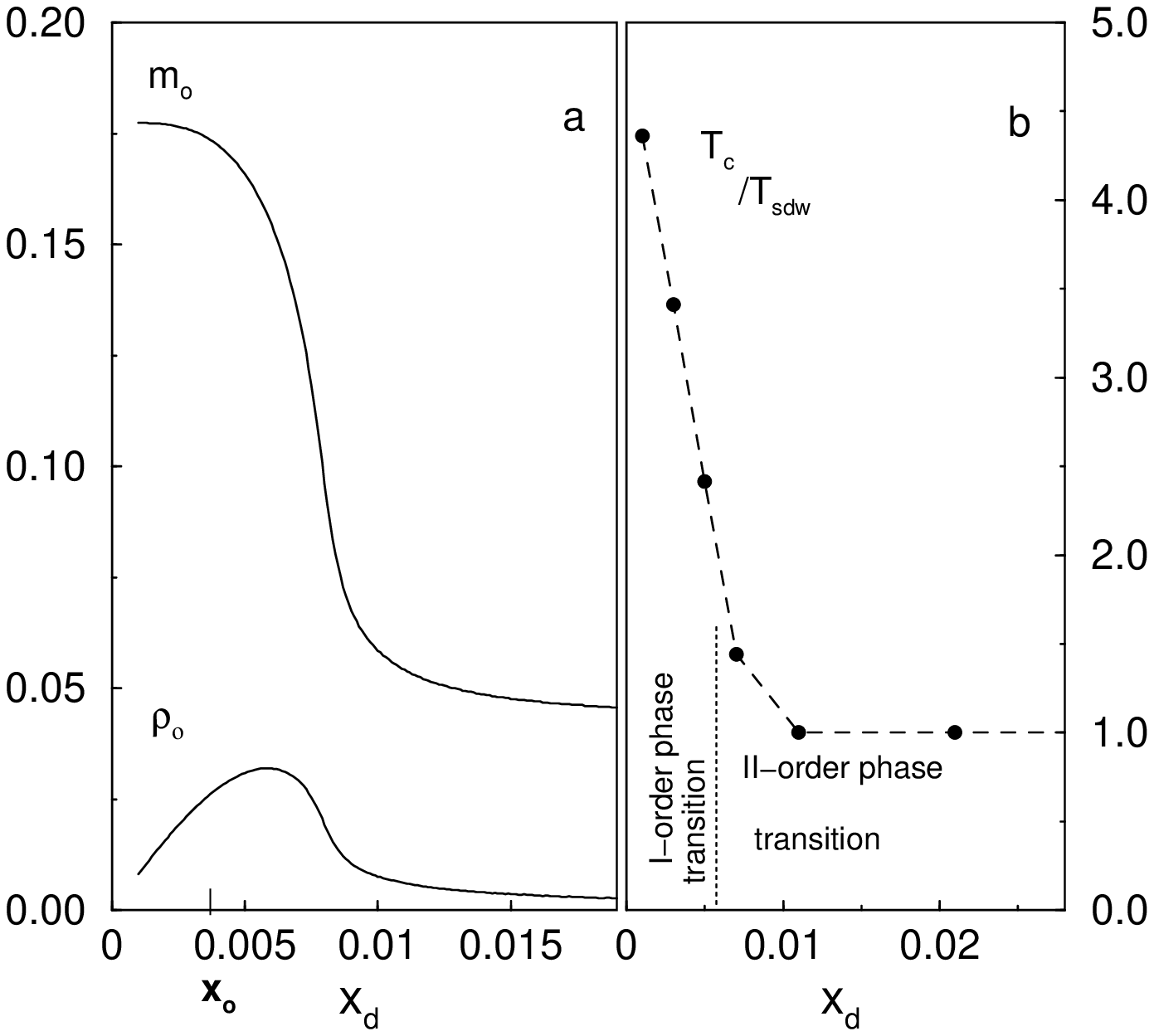}
       }
  \hss}
 }
\vspace{14cm}

\caption{}

\label{fig6}
\end{figure}

\newpage

\begin{figure}
 \vbox to 4.0cm {\vss\hbox to -5.0cm
 {\hss\
       {\includegraphics{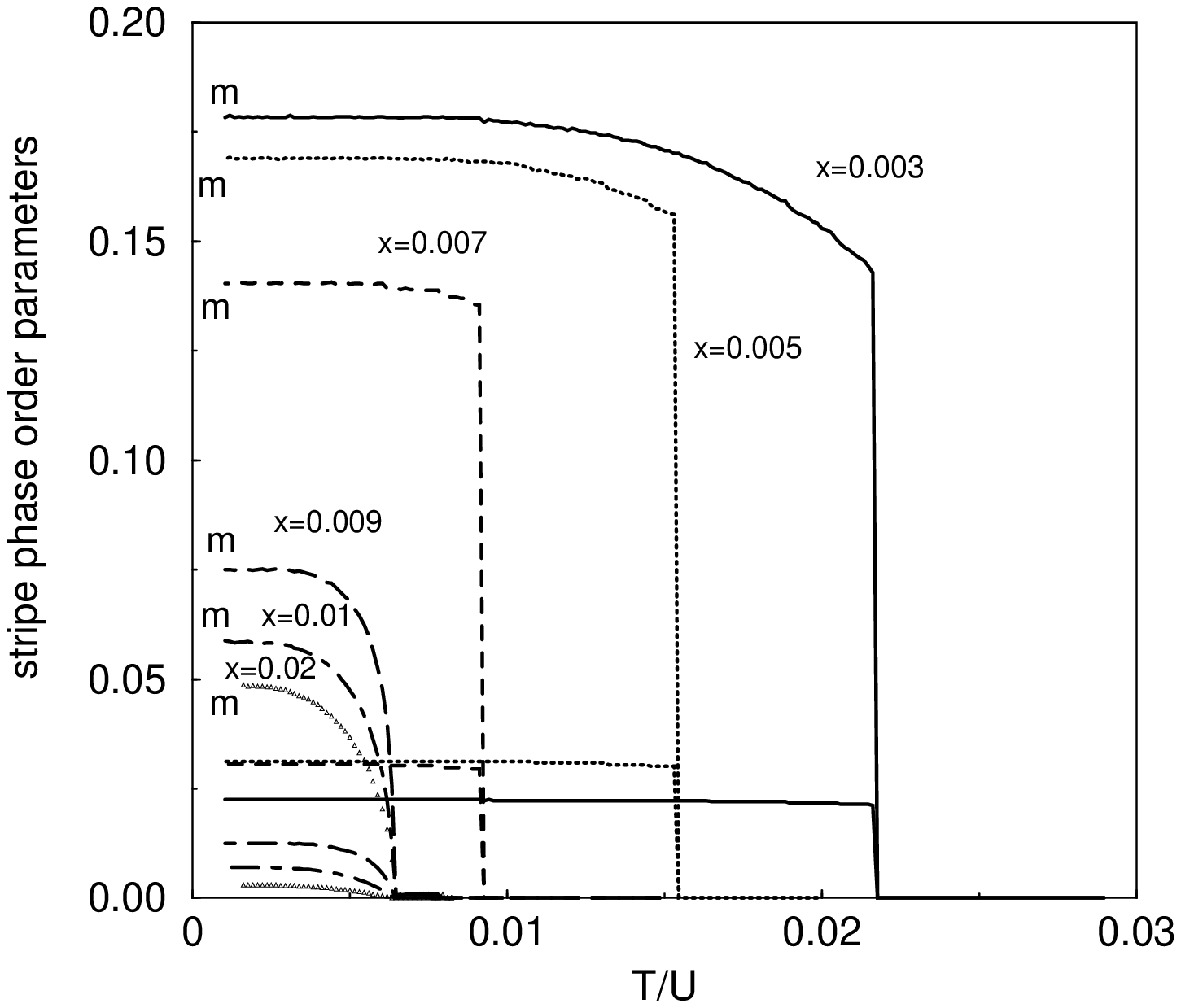}
       }
  \hss}
 }
\vspace{14cm}

\caption{}

\label{fig7}
\end{figure}

\newpage

\begin{figure}
 \vbox to 4.0cm {\vss\hbox to -5.0cm
 {\hss\
       {\includegraphics{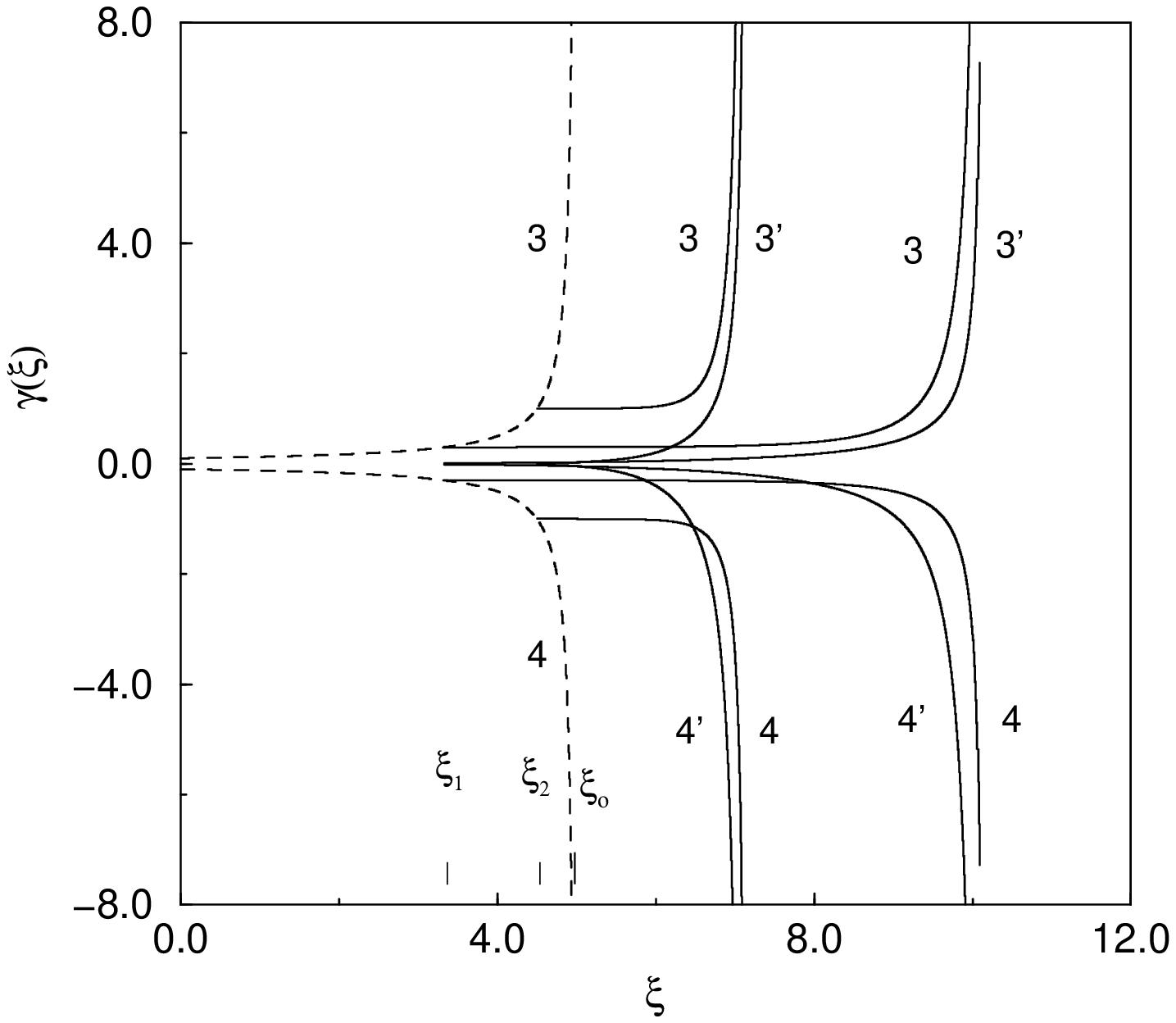}
       }
  \hss}
 }
\vspace{14cm}

\caption{}

\label{fig8}
\end{figure}

\newpage

\begin{figure}
 \vbox to 4.0cm {\vss\hbox to -5.0cm
 {\hss\
       {\includegraphics{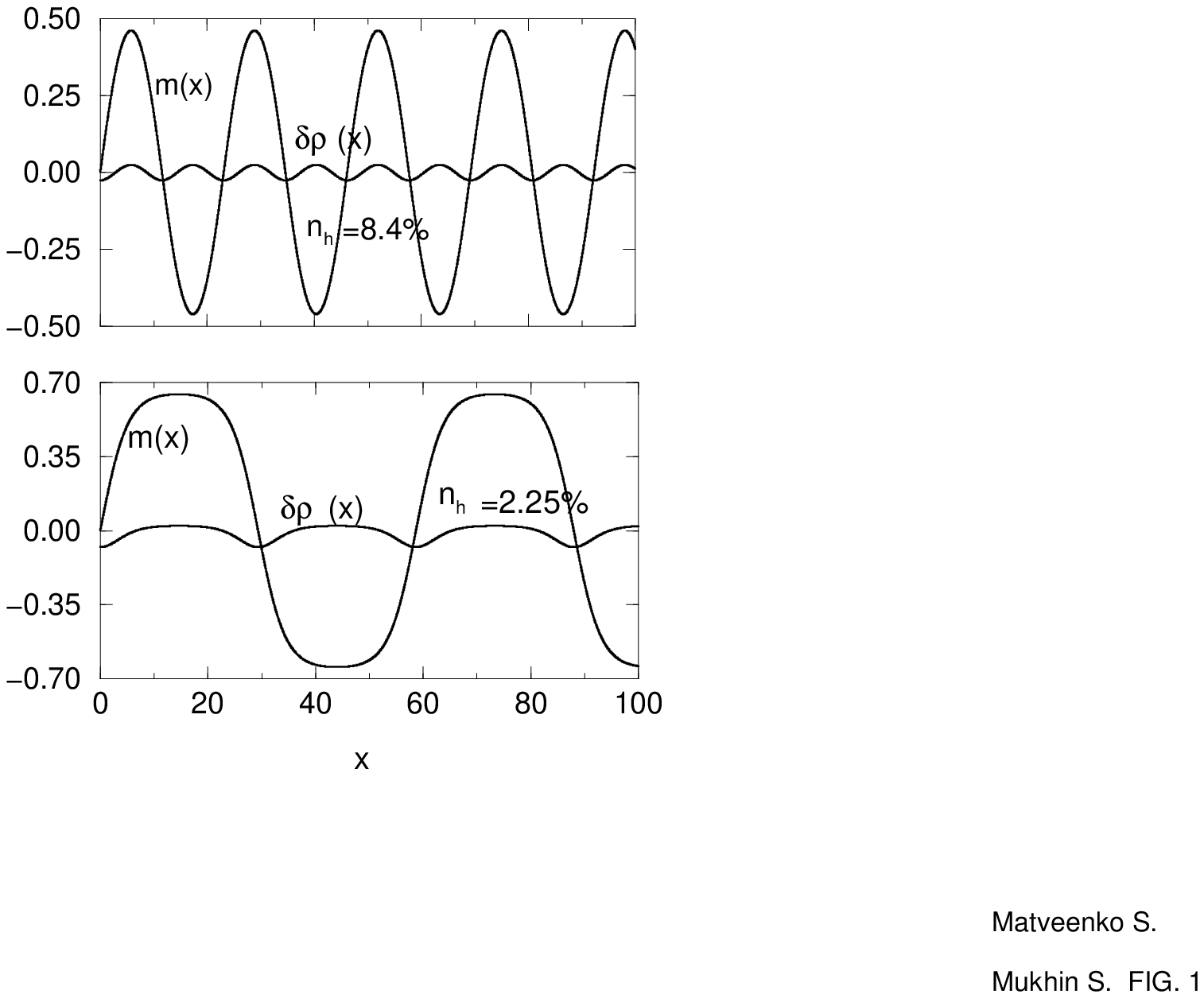}
       }
  \hss}
 }
\vspace{14cm}

\caption{}

\label{fig9}
\end{figure}

\newpage

%


%

\begin{figure}
\epsfxsize=4in
\centerline{\epsfbox{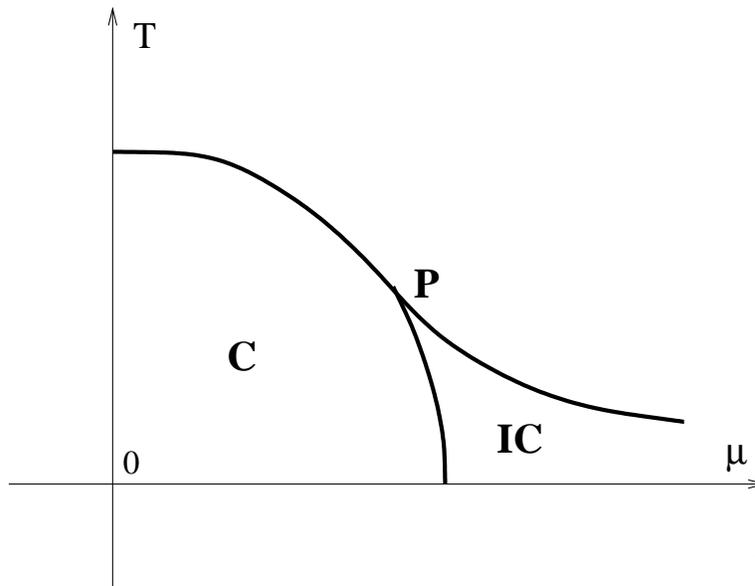}}
\caption[]{The qualitative phase diagram: the phase transition between stripe and
antiferromagnet state is first order, other transition lines are second order.
All three lines of phase transitions are intersected at the Lifshitz point P.}
\label{fig10}
\end{figure}



\begin{references}
\bibitem{rem} Formally, a leading harmonic approximation prevents one from keeping
powers of $z^{n}\ll 1$ with $n$ higher than 1 in the Bessel functions, but we use
symbols $J_n$ to
make more transparent derivation of Eq. (\ref{f}) from Eq. (\ref{Bessel}).
\bibitem{coul}  Long-range Coulomb interaction would certainly influence the total
energy balance of the stripe phase. This paper is focused on the major role of the
on-site repulsion in the ``selection'' of the coupled SDW and CDW modes.
\end{references}
\end{document}